\documentclass[aps,prc,reprint,groupedaddress]{revtex4-1}
\usepackage[utf8]{inputenc}
\usepackage{amsmath}
\usepackage{amsfonts}
\usepackage{amssymb}
\usepackage{graphicx}
\usepackage{slashed}
\begin{document}
	\title{Basis light front quantization for the charged light mesons with color singlet Nambu--Jona-Lasinio interactions}
	\author{Shaoyang Jia}
	\email[]{sjia@iastate.edu}
	\author{James P. Vary}
	\email[]{jvary@iastate.edu}
	\affiliation{Department of Physics and Astronomy, Iowa State University, Ames, Iowa 50011, USA}
	\date{\today}
	\begin{abstract}
		We apply the basis light front quantization (BLFQ) approach to describe the valence structures of the charged light meson ground states. Specifically, the light front wavefunctions of $\pi^\pm$, $\rho^\pm$, $K^\pm$, and $K^{*\pm}$ are obtained as the eigenvectors of the light front effective Hamiltonians with confinement potentials supplemented by the color singlet Nambu--Jona-Lasinio (NJL) interactions. We adjust our model such that the spectrum of these ground states and the charge radii of the pseudoscalar states agree with experimental results. We present the elastic form factors and parton distribution amplitudes (PDAs) as illustrations of the internal structures of the pseudoscalar pions and kaons in terms of valence quarks. 
	\end{abstract}
	\maketitle
\section{Introduction\label{sc:introduction}}
Light mesons provide testing grounds for nonperturbative approaches to quantum chromodynamics (QCD), as explaining their structures in terms of quarks and gluons requires formulating the strong interaction beyond the perturbative expansions. While the underlying symmetry of QCD is the $\mathrm{SU}(3)$ local color gauge symmetry, in the limit of vanishing quark mass there exist global chiral symmetries in the Lagrangian. However, these chiral symmetries are broken by nonzero quark masses and by the dynamics of QCD. With only quarks as the effective degrees of freedom, the Nambu--Jona-Lasinio (NJL) model maintains the local chiral symmetry in the Lagrangian while allowing for the dynamical breaking of such symmetry~\cite{Klimt:1989pm,Vogl:1989ea,Vogl:1991qt,Klevansky:1992qe}. Therefore the NJL interactions are natural candidates for the effective dynamics of quarks inside the light mesons. With equal time quantization, light meson structures have been solved using the NJL model within the framework of Bethe-Salpeter equations (BSEs) in Refs.~\cite{Carrillo-Serrano:2015uca,Hutauruk:2016sug,Hutauruk:2018zfk}. With the NJL interactions, one could also solve for the baryon systems~\cite{Cloet:2014rja,Carrillo-Serrano:2014zta,Carrillo-Serrano:2016igi}. While with QCD interactions, various approaches have been applied to solve for the structures of meson systems, including BSE~\cite{Maris:1999bh,Maris:1999ta,Maris:1999nt,Maris:2000sk,Maris:2002mz,Bender:2002as,Bhagwat:2006pu,Cloet:2013jya,Shi:2015esa,Chen:2016sno,Shi:2018mcb,Shi:2018zqd,Williams:2009ce,Luecker:2009bs,Fischer:2014xha,Heupel:2014ina,Fischer:2014cfa,Williams:2015cvx,Weil:2017knt,Raya:2017ggu,Mojica:2017tvh}, lattice QCD~\cite{Aoki:2016frl,Beane:2011zm,Hoelbling:2011kk,Durr:2010hr,Scholz:2009yz,Bernard:2001fz,Bernard:2002pc,Beane:2006kx}, and other approaches~\cite{Yabusaki:2015dca,deMelo:2016kiu,Mello:2017mor,Ahmady:2018muv}.

In the light front quantization framework, quantization conditions for fields are specified at equal light front time $x^+=x^0+x^3$. Within this framework, the basis light front quantization (BLFQ) is a Hamiltonian approach for solving bound state problems. Specifically, the light front wavefunctions, expanded in terms of orthonormal basis functions, are obtained as eigenfunctions of the light front Hamiltonian. The Hamiltonian of the system under investigation then takes the form of a matrix in the representation of these orthonormal functions. A significant advantage of the light front Hamiltonian approach is the facility for evaluating observables using the resulting light front wavefunctions as we illustrate for the charged light mesons in the present work. 

Earlier applications of the BLFQ were developed for the positronium system with a discretized longitudinal momentum basis and the two-dimensional (2D) harmonic oscillator basis for the transverse momenta~\cite{Maris:2013qma,Li:2013cga,Wiecki:2013cba,Wiecki:2014ola,Wiecki:2015xxa,Adhikari:2016idg}. The addition of a longitudinal confinement potential to the effective Hamiltonian allowed the expansion of the longitudinal momentum dependence of the light front wavefunction in terms of square-integrable functions. Combining the transverse and longitudinal confinements with the one-gluon exchange interaction was subsequently applied to the valence structures of heavy quarkonium~\cite{Li:2015zda,Li:2016wwu,Li:2017mlw,Leitao:2017esb,Li:2017uug,Li:2018uif,Adhikari:2018umb}. Meanwhile, further developments within the BLFQ approach are being made for a number of applications in hadron physics~\cite{Vary:2016ccz,Vary:2016emi,Tang:2018myz}.

In this work we describe the structure of charged light mesons in terms of the light front wavefunctions for the valence quarks. By subsuming self-energy contributions to the meson structure from sea quarks and gluons, these valence quarks resemble the constituent quarks. The parameter space of our BLFQ-NJL model is spanned by the quark masses, the confining strengths, and the coupling constants of the NJL interactions. This parameter space is constrained to reproduce the mass spectrum for the $\pi^\pm,~\rho^\pm,~K^\pm$, and $K^{*\pm}$ ground states summarized by the Particle Data Group (PDG)~\cite{Tanabashi:2018oca}. Aside from the mass spectrum, we also consider the comparison of $\pi^+$ and $K^+$ elastic form factors with available experimental data. The charge radii extracted from these form factors are then used to fix the remaining scales of our model. We also present decay constants and parton distribution amplitudes (PDAs) of $\pi^+$, $\rho^+$, $K^+$, and $K^{*+}$. 

Experimental results on the elastic form factor for the pion are available in Refs.~\cite{Amendolia:1984nz,Amendolia:1986wj,Horn:2006tm,Tadevosyan:2007yd,Huber:2008id}. Measurements of the elastic form factor for the kaon can be found in Refs.~\cite{Dally:1980dj,Amendolia:1986ui,Carmignotto:2018uqj}, with data from the JLab $12~\mathrm{GeV}$ experiment expected \cite{Horn:2017naf}. References~\cite{Gao:2017mmp,Horn:2016rip} and references therein provide additional discussions on these elastic form factors. For the pion PDA, Refs.~\cite{Aitala:2000hb,Bakulev:2003cs} provide experimental results. A selection of theoretical calculations of the pion and the kaon PDAs can be found in Refs.~\cite{Brodsky:2008pg,Cloet:2013tta,Segovia:2013eca,Gao:2014bca,Shi:2014uwa,Shi:2015esa}. 

This article is organized as follows. Section~\ref{sc:introduction} gives the introduction. Section~\ref{sc:NJL} briefly introduces the Lagrangian of the color singlet NJL model. Section~\ref{sc:BLFQ} describes the BLFQ framework and how to evaluate the matrix elements of the NJL interactions within this framework. We present our results for decay constants, charge radii, elastic form factors and parton distribution amplitudes in Sec.~\ref{sc:results}. The summary and concluding remarks are given in Sec.~\ref{sc:summary}.
\section{Color singlet Nambu--Jona-Lasinio interactions\label{sc:NJL}}
With only quarks as the explicit degrees of freedom, the Nambu--Jona-Lasinio model is constructed through a Lagrangian that preserves global chiral symmetries. Therefore within this model of low-energy QCD, dynamics due to gluon-quark coupling and gluon self-couplings are absorbed into local fermion self-interactions. The Lagrangian of color singlet four-fermion interactions in the three-flavor NJL model is given by Ref.~\cite{Klimt:1989pm} as 
\begin{align}
\mathcal{L}_{\mathrm{NJL},\mathrm{SU}(3)}^{(4)} & =\overline{\psi}(i\slashed{\partial}-m)\psi \nonumber\\
& \quad  +G_{\pi}\sum_{i=0}^{8}\left[\left(\overline{\psi}\lambda^i\psi\right)^2+\left(\overline{\psi}i\gamma_5\lambda^i\psi\right)^2\right]\nonumber\\
&\quad  -G_{\rho}\sum_{i=0}^{8}\left[\left(\overline{\psi}\gamma_\mu\lambda^i\psi\right)^2+\left(\overline{\psi}\gamma_\mu\gamma_5\lambda^i\psi\right)^2 \right]\nonumber\\
&\quad  -G_{\mathrm{V}}\left(\overline{\psi}\gamma_\mu\psi\right)^2-G_{\mathrm{A}}\left(\overline{\psi}\gamma_\mu\gamma_5\psi\right)^2.\label{eq:NJL_SU3}
\end{align}
Here $\psi=(\mathrm{u},\mathrm{d},\mathrm{s})^{\mathrm{T}}$ with $\mathrm{u}$, $\mathrm{d}$, and $\mathrm{s}$ representing the up, down and strange quark Dirac spinor fields respectively. The letter ``T'' on the superscript stands for the matrix transpose. The $\lambda^i$ are the Gell-Mann matrices in the flavor space. The coefficients $G_{\pi}$, $G_{\rho}$, $G_{\mathrm{V}}$, and $G_{\mathrm{A}}$ are independent coupling constants of the theory. In this article, we consider color singlet NJL interactions in the form of Eq.~\eqref{eq:NJL_SU3} only, although color octet interactions are also available in Ref.~\cite{Klimt:1989pm}.

Within the NJL Lagrangian given by Eq.\eqref{eq:NJL_SU3}, local chiral symmetries are explicitly broken only by the nonvanishing quark mass, while dynamical chiral symmetry breaking happens at the level of Green's functions. The global $\mathrm{U}(1)$ axial symmetry is still preserved by the interactions in Eq.~\eqref{eq:NJL_SU3}. However, this axial symmetry is broken in QCD by field theory effects. Determinant terms can be introduced to account for such effects in the NJL model~\cite{Klimt:1989pm}. Explicitly, we have the following term in additional to Eq.~\eqref{eq:NJL_SU3}:
\begin{equation}
\mathcal{L}_{\det}=G_{\mathrm{D}}\left[\det\,\overline{\psi}(1+\gamma_5)\psi+ \det\,\overline{\psi}(1-\gamma_5)\psi\right],\label{eq:NJL_det}
\end{equation}
where the determinants are taken in the flavor space, resulting in six-fermion interactions. Notice that aside from the kinematic term, interactions in Eqs.~(\ref{eq:NJL_SU3},~\ref{eq:NJL_det}) are all local contact interactions.

In the scenario where only the up quarks and the down quarks are active, determinant terms in Eq.~\eqref{eq:NJL_det} are reduced to four-fermion interactions. Explicitly, we have
\begin{align}
\mathcal{L}_{\mathrm{det}}&=2G_{\mathrm{D}}\big\{\overline{\mathrm{u}}\mathrm{u}\,\overline{\mathrm{d}}\mathrm{d}+\overline{\mathrm{u}}\gamma_5 \mathrm{u}\overline{\mathrm{d}}\gamma_5 \mathrm{d}-\overline{\mathrm{u}}\mathrm{d}\,\overline{\mathrm{d}}\mathrm{u}-\overline{\mathrm{u}}\gamma_5 \mathrm{d}\,\overline{\mathrm{d}}\gamma_5\mathrm{u} \big\}.
\end{align}
In this case of only two light quarks, one version of the Lagrangian for the NJL model is given by
\begin{align}
\mathcal{L}_{\mathrm{NJL},\mathrm{SU}(2)}&=\overline{\psi}(i\slashed{\partial}-m)\psi +\frac{G_{\pi}}{2}\left[(\overline{\psi}\psi)^2-\left(\overline{\psi}\gamma_5\overrightarrow{\tau}\psi \right)^2 \right]\nonumber\\
&\quad -\frac{G_{\rho}}{2}\left[(\overline{\psi}\gamma_\mu\overrightarrow{\tau}\psi)^2-\left(\overline{\psi}\gamma_\mu\gamma_5\overrightarrow{\tau}\psi \right)^2 \right]\nonumber\\
&\quad -G_{\mathrm{V}}\left(\overline{\psi}\gamma_\mu\psi \right)^2-G_{\mathrm{A}}\left(\overline{\psi}\gamma_\mu\gamma_5\psi \right)^2,\label{eq:NJL_SU2}
\end{align}
which is consistent with the three-flavor Lagrangian in Eq.~\eqref{eq:NJL_SU3} when determinant terms defined by Eq.~\eqref{eq:NJL_det} are added. After setting $G_A=0$, Eq.~\eqref{eq:NJL_SU2} is reduced to the NJL Lagrangian in Ref.~\cite{Carrillo-Serrano:2015uca}.

\section{Basis light front quantization in the meson valence quark Fock sector\label{sc:BLFQ}}
\subsection{Basis representation for the meson valence quark Fock sector\label{ss:basis_rep}}
Within the framework of light front quantization, structures of bound states are embedded in the light front wavefunctions $\vert \Psi\rangle$. The light front wavefunction is solved from the light front Schr\"{o}dinger equation 
\begin{equation}
H_{\mathrm{eff}}\vert \Psi\rangle=M^2\vert \Psi\rangle,\label{eq:LF_Schrodinger}
\end{equation}
where $H_{\mathrm{eff}}$ is the effective Hamiltonian of the system. For simplicity, our model only concerns the valence quark Fock sector of the mesons, leaving contributions to the hadron structures from sea quarks and gluons implicit. We only need to construct the valence Fock sector wavefunctions for the positively charged mesons, leaving properties of negatively charge mesons obtainable through the charge conjugation. 

Explicitly, when the Fock space of the hadron wavefunction is truncated to the valence quark and antiquark, the light front eigenstate for the positively charged meson is given by
\begin{align}
&\quad \big\vert\Psi_{\mathrm{meson}}(P^+,\overrightarrow{P}^\perp,J,m_J)\big\rangle \nonumber\\
&=\sum_{r,s}\int_{0}^{+\infty}\dfrac{dk^+}{4\pi k^+}\int \dfrac{d\overrightarrow{k}^\perp}{(2\pi)^2}\int_{0}^{+\infty}\dfrac{dp^+}{4\pi p^+}\int \dfrac{d\overrightarrow{p}^\perp}{(2\pi)^2}\nonumber\\
&\quad \times 4\pi P^+\delta(k^++p^+-P^+)(2\pi)^2\delta\left(\overrightarrow{k}^\perp+\overrightarrow{p}^\perp-\overrightarrow{P}^\perp \right)\nonumber\\
&\quad\hspace{2.25cm}  \times \Psi_{rs}(k,p;P,J,m_J)b_r^\dagger(k)d_{s}^\dagger(p)|0\rangle\label{eq:Psi_meson_qqbar_ori}\\
& =\sum_{r,s}\int_{0}^{1}\dfrac{dx}{4\pi x(1-x)}\int\dfrac{d\overrightarrow{\kappa}^\perp}{(2\pi)^2}\,\psi_{rs}(x,\overrightarrow{\kappa}^\perp)\nonumber\\
&
\quad \times b_r^\dagger(xP^+,\overrightarrow{\kappa}^\perp+x\overrightarrow{P}^\perp)\nonumber\\
&\quad \hspace{1.2cm}\times d_s^\dagger((1-x)P^+,-\overrightarrow{\kappa}^\perp+(1-x)\overrightarrow{P}^\perp)|0\rangle,\label{eq:Psi_meson_qqbar}
\end{align}
where $P=k+p$ is the total momentum of the meson, $x=k^+/P^+$ is the longitudinal momentum fraction carried by the valence quark, and $\overrightarrow{\kappa}^\perp=\overrightarrow{k}^\perp-x\overrightarrow{P}^\perp$ is the relative transverse momentum. In this limited Fock space, the meson has fixed total angular momentum projection $m_J$. The total angular momentum $J$ is dynamical in light front quantization, the consevation of which is expected if one does not truncate the Fock space expansion. We find it is approximately conserved in our previous BLFQ works for low-lying states~\cite{Wiecki:2013cba,Wiecki:2014ola,Wiecki:2015xxa,Li:2015zda,Li:2016wwu,Li:2017mlw,Leitao:2017esb,Li:2017uug,Li:2018uif,Adhikari:2018umb,Vary:2016ccz,Vary:2016emi} as well as in this work so we include it as a label for our solutions. In Eq.~\eqref{eq:Psi_meson_qqbar_ori}, the creation operator $d^\dagger_{s}(p)$ creates an antiquark with spin $s$ and light front $3$-momentum $p$ from the vacuum, while the operator $b^\dagger_{r}(k)$ creates a quark with spin $r$ and momentum $k$. In Eq.~\eqref{eq:Psi_meson_qqbar}, momenta are written in terms of longitudinal momentum fractions and the relative transverse momentum, such that the conservation of the total light front $3$-momentum $P$ is ensured. The function $\psi_{rs}(x,\overrightarrow{\kappa}^\perp)$ is recognized as the valence Fock sector light front wavefunction in momentum space for the meson, with its normalization defined by
\begin{equation}
\sum_{r,s}\int_{0}^{1}\dfrac{dx}{4\pi x(1-x)}\int\dfrac{d\overrightarrow{\kappa}^\perp}{(2\pi)^2}\psi^*_{rs}(x,\overrightarrow{\kappa}^\perp)\psi_{rs}(x,\overrightarrow{\kappa}^\perp)=1.\label{eq:normalization_psi_rs_xkappa}
\end{equation}

In order to solve the light front wavefunction $\psi_{rs}(x,\overrightarrow{\kappa}^\perp)$ from Eq.~\eqref{eq:LF_Schrodinger}, we formally decompose the effective Hamiltonian into a two-body term and an interaction term. Explicitly, we write down
\begin{equation}
H_{\mathrm{eff}}=H_0+H^{\mathrm{eff}}_{\mathrm{int}},\label{eq:decompose_Heff}
\end{equation}
where $H_0$ contains the kinematic terms and the two-body confinement potentials, leaving $H^{\mathrm{eff}}_{\mathrm{int}}$ with the remaining interaction terms. Following Refs.~\cite{Li:2015zda,Li:2017mlw}, in the valence Fock sector $H_0$ is given by
\begin{align}
H_0& =\dfrac{\overrightarrow{\kappa}_\perp^2+\mathbf{m}^2}{x}+\dfrac{\overrightarrow{\kappa}_\perp^2+\overline{\mathbf{m}}^2}{1-x}+\kappa^4x(1-x)\overrightarrow{r}_\perp^2\nonumber\\
&\quad -\dfrac{\kappa^4}{(\mathbf{m}+\overline{\mathbf{m}})^2}\partial_xx(1-x)\partial_x,\label{eq:H0_def}
\end{align}
where the first two terms are the kinematic energy of the quark and the antiquark, the third term is the transverse confinement potential, and the last term is the longitudinal confinement potential. Here $\mathbf{m}$ and $\mathbf{\overline{m}}$ are the quark mass and the antiquark mass respectively. Notice that the vector $\overrightarrow{\kappa}_\perp$ stands for the relative transverse momentum of the valence quarks while $\overrightarrow{r}_\perp$ is the conjugate variable of $\overrightarrow{\kappa}_\perp$. The scalar $\kappa$ is the confining strength, which is unrelated to the modulus of $\overrightarrow{\kappa}_\perp$. 

Following Refs.~\cite{Li:2015zda,Li:2017mlw}, we then choose the expansion of the light front wavefunction for the valence quarks:
\begin{align}
&\quad \psi_{rs}(x,\overrightarrow{\kappa}^\perp)\nonumber\\
&=\sum_{nml}\psi(n,m,l,r,s)\,\phi_{nm}\left(\dfrac{\overrightarrow{\kappa}^\perp}{\sqrt{x(1-x)}}\right)\chi_l(x),\label{eq:psi_rs_basis_expansions}
\end{align}
where $\phi_{nm}$ is the 2D harmonic oscillator function, and $\chi_l$ is the longitudinal basis function. Explicitly, $\phi_{nm}$ is defined as
\begin{align}
\phi_{nm}\left(\overrightarrow{q}^\perp;b \right) & =\dfrac{1}{b}\sqrt{\dfrac{4\pi n!}{(n+|m|)!}} \left(\dfrac{\vert\overrightarrow{q}^\perp\vert}{b}\right)^{|m|} \exp\left(-\dfrac{\overrightarrow{q}^{\perp 2}}{2b^2}\right)\nonumber\\
&\quad \times \,L_n^{|m|} \left(\dfrac{\overrightarrow{q}^{\perp 2}}{b^2}\right)\,e^{im\varphi},\label{eq:def_phi_nm}
\end{align}
with $\tan(\varphi)=q^2/q^1$ and $L_n^{|m|}$ being the associated Laguerre function. Meanwhile, $\chi_l(x)$ is given by 
\begin{align}
&\quad \chi_l(x;\alpha,\beta)\nonumber\\
&= \sqrt{4\pi(2l+\alpha+\beta+1)}\sqrt{\dfrac{\Gamma(l+1)\Gamma(l+\alpha+\beta+1)}{\Gamma(l+\alpha+1)\Gamma(l+\beta+1)}}\nonumber\\
&\quad \hspace{2cm}\times x^{\beta/2}(1-x)^{\alpha/2}\,P_l^{(\alpha,\beta)}(2x-1),\label{eq:def_chi_l}
\end{align}
with $P_{l}^{(\alpha,\beta)}(z)$ being the Jacobi polynomial and 
\begin{subequations}\label{eq:def_alpha_beta}
	\begin{align}
	\alpha& =2\overline{\mathbf{m}}(\mathbf{m}+\overline{\mathbf{m}})/\kappa^2,\\
	\beta&=2\mathbf{m}(\mathbf{m}+\overline{\mathbf{m}})/\kappa^2.
	\end{align}
\end{subequations}
Additionally, in terms of the basis expansion defined by Eq.~\eqref{eq:psi_rs_basis_expansions}, the normalization condition specified by Eq.~\eqref{eq:normalization_psi_rs_xkappa} becomes
\begin{equation}
\sum_{nmlrs}\psi^*(n,m,l,r,s)\psi(n,m,l,r,s)=1.\label{eq:normalization_basis_expansion}
\end{equation} 

In this article, the default choice of the scale parameter $b$ in Eq.~\eqref{eq:def_phi_nm} is identical to the confining strength $\kappa$ in Eq.~\eqref{eq:H0_def}. Then with the definitions of the basis functions given by Eqs.~(\ref{eq:def_phi_nm},~\ref{eq:def_chi_l}), the $H_0$ term in Eq.~\eqref{eq:H0_def} is diagonal with respect to elements of the basis expansion. Explicitly using Eq.~\eqref{eq:Psi_meson_qqbar_ori}, one can show that
\begin{align}
&\quad \bigg\langle \Psi'_{\mathrm{meson}}\left(P'^+,\overrightarrow{P}'^\perp\right)\bigg\vert H_0 \bigg\vert\Psi_{\mathrm{meson}}\left(P^+,\overrightarrow{P}^\perp\right)\bigg\rangle\nonumber\\
& =4\pi P^+\delta\left(P'^+-P^+\right)(2\pi)^2\delta\left(\overrightarrow{P}'^\perp-\overrightarrow{P}^\perp \right)\nonumber\\
&\quad \times\sum_{r',s'}\sum_{r,s}\delta_{r'r}\delta_{s's}\int_{0}^{1}\dfrac{dx'}{4\pi x'(1-x')} \int\dfrac{d\overrightarrow{\kappa}'^\perp}{(2\pi)^2}\nonumber\\
& \quad\quad \times\int_{0}^{1}\dfrac{dx}{4\pi x(1-x)}\int\dfrac{d\overrightarrow{\kappa}^\perp}{(2\pi)^2}\,\psi'^*_{r's'}(x',\overrightarrow{\kappa}'^\perp)\,\nonumber\\
&\quad \hspace{4.5cm}\times H_0\,\psi_{rs}(x,\overrightarrow{\kappa}^\perp).\label{eq:psibar_H0_psi}
\end{align}
Here $\vert\Psi'\rangle$ and $\vert \Psi \rangle$ are independent kets with the same $J$ and $m_J$ which have been suppressed to simplify the notation. Therefore Eq.~\eqref{eq:psibar_H0_psi} allows the matrix element of $H_0$ to be calculated for arbitrary wavefunctions $\psi'_{r's'}(x',\overrightarrow{\kappa}^\perp)$ and $\psi_{rs}(x,\overrightarrow{\kappa}^\perp)$. This allows any basis representation, including Eq.~\eqref{eq:psi_rs_basis_expansions}, to be applied for the eigenvalue problem of the effective Hamiltonian. 

Specifically, with the basis expansion of $\psi_{rs}(x,\overrightarrow{\kappa}^\perp)$ in the form of Eq.~\eqref{eq:psi_rs_basis_expansions}, Eq.~\eqref{eq:psibar_H0_psi} becomes
\begin{align}
& \quad \bigg\langle \Psi'_{\mathrm{meson}}\left(P'^+,\overrightarrow{P}'^\perp\right)\bigg\vert H_0\nonumber \bigg\vert\Psi_{\mathrm{meson}}\left(P^+,\overrightarrow{P}^\perp\right)\bigg\rangle\nonumber\\
& =4\pi P^+\delta\left(P'^+-P^+\right)(2\pi)^2\delta\left(\overrightarrow{P}'^\perp-\overrightarrow{P}^\perp \right)\nonumber\\
& \quad\times\sum_{n'm'l's1's2'}\sum_{nmls1s2}\delta_{n'n}\delta_{m'm}\delta_{l'l}\delta_{s1's1}\delta_{s2's2}\nonumber\\
&\quad \times\Lambda_0(n,m,l)\psi'^{\,*}(n',m',l',s_1',s_2')\psi(n,m,l,s_1,s_2).\label{eq:psibar_H0_psi_basis_rep}
\end{align}
When the light front wavefunction is given by one basis function from Eq.~\eqref{eq:psi_rs_basis_expansions} such that $\psi'(n',m',l',s'_1,s'_2)=\delta_{n'N'}\delta_{m'M'}\delta_{l'L'}\delta_{s1'S1'}\delta_{s2'S2'}$ and $\psi(n,m,l,s_1,s_2)=\delta_{nN}\delta_{mM}\delta_{lL}\delta_{s1S1}\delta_{s2S2}$, Eq.~\eqref{eq:psibar_H0_psi_basis_rep} is apparently a diagonal matrix: 
\begin{align}
&\quad \langle n',m',l',s'_1,s'_2\vert H_0\vert n,m,l,s_1,s_2\rangle\nonumber\\
& =\Lambda_0(n,m,l)\delta_{n'n}\delta_{m'm}\delta_{s1's1}\delta_{s2's2},
\end{align}
with its eigenvalue given by
\begin{align}
&\quad \Lambda_0(n,m,l;\mathbf{m},\overline{\mathbf{m}},\kappa)\nonumber\\
&=(\mathbf{m}+\overline{\mathbf{m}})^2+2\kappa^2(2n+|m|+l+3/2)\nonumber\\
&\quad \hspace{3cm} +\dfrac{\kappa^4}{(\mathbf{m}+\overline{\mathbf{m}})^2}l(l+1).\label{eq:diagonal_matrix_elements}
\end{align}
\subsection{Matrix elements of the NJL interactions}
The light front Hamiltonian $P^-$ is obtained by the Legendre transform of the corresponding Lagrangian after we eliminate the constrained field components, which usually leads to instantaneous interactions. The effective Hamiltonian in Eq.~\eqref{eq:LF_Schrodinger} is then related to the light front Hamiltonian by ${H_{\mathrm{eff}}=P^+P^- - \overrightarrow{P}^{\perp 2}}$, where $P^+$ and $\overrightarrow{P}^\perp$ are the conserved total momenta. While introducing the NJL dynamics into the BLFQ effective Hamiltonian, we ignore instantaneous interactions due to the NJL interactions for simplicity. Therefore we only consider the contribution to the light front effective Hamiltonian from the NJL interactions directly from the Legendre transform, which is simply the interaction term in the NJL Lagrangian multiplied by $-P^+$. We then take the obtained Hamiltonian term as the interaction term $H_{\mathrm{int}}^{\mathrm{eff}}$ in Eq.~\eqref{eq:decompose_Heff}. 

Specifically for the $\pi^+$ and $\rho^+$ mesons, the flavor structure of their valence quarks is $\mathrm{u\overline{d}}$. If the $G_{\pi}$ term in Eq.~\eqref{eq:NJL_SU2} is kept as the only nonvanishing NJL interaction, after the Legendre transform we obtain
\begin{align}
&\quad H_{\mathrm{NJL},\pi}^{\mathrm{eff}}\nonumber\\
&=\int dx^-\int d\overrightarrow{x}^\perp\,\left(-\dfrac{G_\pi P^+}{2}\right)\left[\left(\overline{\psi}\psi \right)^2+\left(\overline{\psi}i\gamma_5\overrightarrow{\tau}\psi \right)^2 \right]\label{eq:H_eff_NJL_pi_ori}
\end{align}
as the $H_{\mathrm{int}}^{\mathrm{eff}}$ term in Eq.~\eqref{eq:decompose_Heff} for the $\pi$-$\rho$ system. Here the fermion field is given by $\psi=(\mathrm{u},\mathrm{d})^\mathrm{T}$, and $\tau^i$ are the Pauli matrices in the flavor space. We do not use the $G_{\rho}$ term, because the role of binding rho mesons is taken by the confinement potentials in Eq.~\eqref{eq:H0_def}. After expanding in the flavor space, Eq.~\eqref{eq:H_eff_NJL_pi_ori} then becomes
\begin{align}
&\quad H_{\mathrm{NJL},\pi}^{\mathrm{eff}}\nonumber\\
& =\int dx^-\int d\overrightarrow{x}^\perp\,\left(-\dfrac{G_\pi P^+}{2}\right)\Big\{2\,\overline{\mathrm{u}}\mathrm{u}\,\overline{\mathrm{d}}\mathrm{d} +2\,\overline{\mathrm{u}}\gamma_5\mathrm{u}\,\overline{\mathrm{d}}\gamma_5 \mathrm{d}\nonumber\\
&\quad -4\,\overline{\mathrm{u}}\gamma_5\mathrm{d}\,\overline{\mathrm{d}}\gamma_5 \mathrm{u} +(\overline{\mathrm{u}}\mathrm{u})^2+(\overline{\mathrm{d}}\mathrm{d})^2-(\overline{\mathrm{u}}\gamma_5 \mathrm{u})^2-(\overline{\mathrm{d}}\gamma_5 \mathrm{d})^2\Big\}.\label{eq:H_eff_NJL_pi}
\end{align}

In terms of creation and annihilation operators,
we explicitly write down the eigenstate for the positively charged mesons with up and anti-down valence quarks as
\begin{align}
&\quad \big\vert\Psi_{\mathrm{meson}+}(P^+,\overrightarrow{P}^\perp)\big\rangle \nonumber\\
& = \sum_{r,s}\int_{0}^{1}\dfrac{dx}{4\pi x(1-x)}\int\dfrac{d\overrightarrow{\kappa}^\perp}{(2\pi)^2}\,\psi_{rs}(x,\overrightarrow{\kappa}^\perp)\nonumber\\
&\quad \times b_{\mathrm{u}r}^\dagger(xP^+,\overrightarrow{\kappa}^\perp+x\overrightarrow{P}^\perp)\nonumber\\
&\quad \hspace{0.5cm}\times d_{\mathrm{d}s}^\dagger((1-x)P^+,-\overrightarrow{\kappa}^\perp+(1-x)\overrightarrow{P}^\perp)|0\rangle.\label{eq:Psi_meson+_qqbar}
\end{align}
Regarding the subscripts of the creation operators in Eq.~\eqref{eq:Psi_meson+_qqbar}, the non-italic letters represent the flavors while the italic letters designate the spins. After ignoring the self-energy contributions, the operator expansion of Eq.~\eqref{eq:H_eff_NJL_pi} relevant to Eq.~\eqref{eq:Psi_meson+_qqbar} is 
\begin{align}
& \quad H_{\mathrm{NJL,\pi}}^{\mathrm{eff}} \nonumber\\
& =\sum_{s1s2s3s4}\int d\underline{k}_1 d\underline{k}_2 d\underline{k}_3 d\underline{k}_4\,4\pi\delta(k_1^++k_2^+-k_3^+-k_4^+)\,\nonumber\\
& \quad \times (2\pi)^2\delta\left(\overrightarrow{k}_1^\perp+\overrightarrow{k}_2^\perp-\overrightarrow{k}_3^\perp-\overrightarrow{k}_4^\perp \right) G_\pi P^+\nonumber\\
&\quad\quad \times \big\{\overline{u}_{\mathrm{u}1}u_{\mathrm{u}4}\,\overline{v}_{\mathrm{d}3}v_{\mathrm{d}2}+\overline{u}_{\mathrm{u}1} \gamma_5 u_{\mathrm{u}4}\,\overline{v}_{\mathrm{d}3}\gamma_5 v_{\mathrm{d}2}\nonumber\\
&\quad \hspace{1.5cm} + 2\, \overline{u}_{\mathrm{u}1}\gamma_5 v_{\mathrm{d}2}\, \overline{v}_{\mathrm{d}3}\gamma_5 u_{\mathrm{u}4} \big\}b_{\mathrm{u}1}^\dagger d_{\mathrm{d}2}^\dagger d_{\mathrm{d}3} b_{\mathrm{u}4},\label{eq:H_NJL_pi_matrix}
\end{align}
where the number subscripts distinguish different fermion spins and momenta while the summation indices are only the spin labels. We use this compact subscripting convention when explicit integration variables permit. Additionally, the momentum space integral measure is defined as 
\begin{equation}
\int d\underline{k}=\int_{0}^{+\infty}\dfrac{dk^+}{4\pi k^+}\int_{-\infty}^{+\infty} \dfrac{dk_1^\perp}{2\pi}\int_{-\infty}^{+\infty} \dfrac{dk_2^\perp}{2\pi}.
\end{equation}

We then evaluate the valence Fock block of the NJL effective Hamiltonian matrix given by Eq.~\eqref{eq:H_NJL_pi_matrix} for the meson wavefunction in Eq.~\eqref{eq:Psi_meson+_qqbar}. Explicitly, we have 
\begin{align}
&\quad  \big\langle\Psi'_{\mathrm{meson}+}(P'^+,\overrightarrow{P}'^\perp)\big\vert H_{\mathrm{NJL},\pi}^{\mathrm{eff}} \big\vert\Psi_{\mathrm{meson}+}(P^+,\overrightarrow{P}^\perp)\big\rangle\nonumber\\
& =4\pi P^+ \delta(P'^+-P^+)(2\pi)^2\delta(\overrightarrow{P}'^\perp-\overrightarrow{P}^\perp)\sum_{s1's2's1s2}\nonumber\\
&\quad \times \int_{0}^{1}\dfrac{dx'}{4\pi x'(1-x')}\int \dfrac{d\overrightarrow{\kappa}'^\perp}{(2\pi)^2}\int_{0}^{1}\dfrac{dx}{4\pi x(1-x)}\int \dfrac{d\overrightarrow{\kappa}^\perp}{(2\pi)^2}\nonumber\\
& \quad \times \psi'^{\,*}_{s1's2'}(x',\kappa'^\perp)\psi_{s1s2}(x,\kappa^\perp)\nonumber\\
&\quad \times G_\pi\, \big\{\overline{u}_{\mathrm{u}s1'}(p_1')u_{\mathrm{u}s1}(p_1)\,\overline{v}_{\mathrm{d}s2}(p_2)v_{\mathrm{d}s2'}(p_2') \nonumber\\
&\quad\quad + \overline{u}_{\mathrm{u}s1'}(p_1')\gamma_5 u_{\mathrm{u}s1}(p_1)\,\overline{v}_{\mathrm{d}s2}(p_2)\gamma_5 v_{\mathrm{d}s2'}(p_2') \nonumber\\
&\quad\quad+ 2\,\overline{u}_{\mathrm{u}s1'}(p_1')\gamma_5 v_{\mathrm{d}s2'}(p_2')\,\overline{v}_{\mathrm{d}s2}(p_2)\gamma_5 u_{\mathrm{u}s1}(p_1) \big\},\label{eq:H_NJL_pi_matrix_wf}
\end{align}
with the momentum dependence of the spinors given by
\begin{subequations}\label{eq:def_relative_momenta}
	\begin{align}
	& p_1'^+ =x'P'^+, && \overrightarrow{p}_1'^\perp=\overrightarrow{\kappa}'^\perp +x'\overrightarrow{P}'^\perp, \\
	& p_2'^+ =(1-x')P'^+, && \overrightarrow{p}_2'^\perp =-\overrightarrow{\kappa}'^\perp+(1-x')\overrightarrow{P}'^\perp,\\
	& p_1^+ =xP^+, && \overrightarrow{p}_1^\perp=\overrightarrow{\kappa}^\perp +x\overrightarrow{P}^\perp, \\
	& p_2^+ =(1-x)P^+, && \overrightarrow{p}_2^\perp =-\overrightarrow{\kappa}^\perp+(1-x)\overrightarrow{P}^\perp.
	\end{align}
\end{subequations}
Spin decompositions of all terms in Eq.~\eqref{eq:H_NJL_pi_matrix_wf} can then be easily calculated using definitions in App.~\ref{ss:spinor_def}. Subsequently for any given combination of the basis functions in Eq.~\eqref{eq:psi_rs_basis_expansions}, because the momentum dependence of the wavefunction is exactly known, we can calculate the corresponding matrix element explicitly. 

Within the basis representation given by Eq.~\eqref{eq:psi_rs_basis_expansions}, let us define the matrix element for the first term in Eq.~\eqref{eq:H_NJL_pi_matrix_wf} through 
\begin{align}
& \quad \big\langle n'm'l's_1's_2'\big\vert \overline{u}_{\mathrm{u}}u_{\mathrm{u}}\,\overline{v}_{\mathrm{d}}v_{\mathrm{d}}\big\vert nmls_1s_2\big\rangle\nonumber\\
& \equiv \int_{0}^{1}\dfrac{dx'}{4\pi}\int_{0}^{1}\dfrac{dx}{4\pi}\chi_{l'}(x')\chi_{l}(x)\,\int\dfrac{d\overrightarrow{q}'^\perp}{(2\pi)^2}\int\dfrac{d\overrightarrow{q}^\perp}{(2\pi)^2}\,\nonumber\\
&\quad \times \phi_{n'm'}^*\left(\overrightarrow{q}'^\perp\right)\phi_{nm}\left(\overrightarrow{q}^\perp\right)\,\overline{u}_{\mathrm{u}s1'}(p_1')u_{\mathrm{u}s1}(p_1)\,\nonumber\\
&\quad\hspace{4cm}\times \overline{v}_{\mathrm{d}s2}(p_2) v_{\mathrm{d}s2'}(p_2'),\label{eq:Heff_scalar_uuvv_matrix_elements}
\end{align}
where the spinor momenta are also given by Eq.~\eqref{eq:def_relative_momenta}. Similarly one can define $\big\langle n'm'l's_1's_2'\big\vert \overline{u}_{\mathrm{u}}\gamma_5 u_{\mathrm{u}}\,\overline{v}_{\mathrm{d}}\gamma_5 v_{\mathrm{d}}\big\vert nmls_1s_2\big\rangle$
and 
$\big\langle n'm'l's_1's_2'\big\vert \overline{u}_{\mathrm{u}}\gamma_5 v_{\mathrm{d}}\,\overline{v}_{\mathrm{d}}\gamma_5 u_{\mathrm{u}}\big\vert nmls_1s_2\big\rangle$. Because the interactions in $H_{\mathrm{int}}^{\mathrm{eff}}$ are all local, the matrix elements in our basis representation can be calculated exactly without resorting to quadrature. The expressions of these matrix elements and details on how to obtain them are given in App.~\ref{ss:NJL_matrix_elements}.

For the $K^+$-$K^{*+}$ systems, we take 
\begin{align}
&\quad H^{\mathrm{eff}}_{\mathrm{NJL},K}\nonumber\\
& =\int dx^-\int d\overrightarrow{x}^\perp\,(-G_{K}P^+)\sum_{i=1}^{8}[(\overline{\psi}\lambda^i\psi)^2-(\overline{\psi}\lambda^i\gamma_5\psi)^2 ],\label{eq:H_eff_NJL_SU_3_ori}
\end{align}
as the $H_{\mathrm{int}}^{\mathrm{eff}}$ term in Eq.~\eqref{eq:decompose_Heff}. Similar to the case of the $\rho$ meson, the binding of $K^*$ meson is addressed by the confining potentials. Meanwhile, the expansion of the Dirac bilinear in the flavor space with the up, down and strange quarks is given by
\begin{align}
\sum_{i=1}^{8}(\overline{\psi}\,\lambda^i\gamma^?\,\psi)^2
& =(\overline{\mathrm{u}}\gamma^?\mathrm{u}+\overline{\mathrm{d}}\gamma^?\mathrm{d})^2+(\overline{\mathrm{u}}\gamma^?\mathrm{d}+\overline{\mathrm{d}}\gamma^?\mathrm{u})^2\nonumber\\
&\quad +2[(\overline{\mathrm{s}}\gamma^?\mathrm{s})^2+2\,\overline{\mathrm{u}}\gamma^?\mathrm{s}\,\overline{\mathrm{s}}\gamma^? \mathrm{u}+2\,\overline{\mathrm{d}}\gamma^?\mathrm{s}\,\overline{\mathrm{s}}\gamma^?\mathrm{d}],\label{eq:SU_3_flavor_decom}
\end{align}
where $(\,\gamma^?\,)^2$ stands for any contractions of gamma matrices. With the assistance of Eq.~\eqref{eq:SU_3_flavor_decom}, Eq.~\eqref{eq:H_eff_NJL_SU_3_ori} becomes
\begin{align}
H^{\mathrm{eff}}_{\mathrm{NJL},K}&=\int dx^-\int d\overrightarrow{x}^\perp\,(-2G_{K}P^+)\big\{\overline{\mathrm{u}}\mathrm{u}\,\overline{\mathrm{d}}\mathrm{d}+\overline{\mathrm{u}}\mathrm{d}\,\overline{\mathrm{d}}\mathrm{u}\nonumber\\
&\quad  +\overline{\mathrm{u}}\mathrm{s}\,\overline{\mathrm{s}}\mathrm{u}-\overline{\mathrm{u}}\gamma_5\mathrm{u}\,\overline{\mathrm{d}}\gamma_5\mathrm{d}-\overline{\mathrm{u}}\gamma_5\mathrm{d}\,\overline{\mathrm{d}}\gamma_5\mathrm{u}-\overline{\mathrm{u}}\gamma_5\mathrm{s}\,\overline{\mathrm{s}}\gamma_5\mathrm{u} \big\}.\label{eq:H_eff_NJL_K_ori}
\end{align}

The wavefunction for positively charged $K$ mesons is given by Eq.~\eqref{eq:Psi_meson+_qqbar} with the flavor subscript of the $d^\dagger_{\mathrm{d}s}$ operator modified into that of $d^\dagger_{\mathrm{s}s}$. After ignoring self-energy contributions, we expand Eq.~\eqref{eq:H_eff_NJL_K_ori} in terms of relevant creation and annihilation operators as
\begin{align}
&\quad H_{\mathrm{NJL,K}}^{\mathrm{eff}} \nonumber\\
& =\sum_{s1s2s3s4}\int d\underline{k}_1 d\underline{k}_2 d\underline{k}_3 d\underline{k}_4\,4\pi\delta(k_1^++k_2^+-k_3^+-k_4^+)\,\nonumber\\
&\quad \times (2\pi)^2\delta\left(\overrightarrow{k}_1^\perp+\overrightarrow{k}_2^\perp-\overrightarrow{k}_3^\perp-\overrightarrow{k}_4^\perp \right) G_K P^+ \nonumber\\
& \quad\times\big\{-2\overline{u}_{\mathrm{u}1} v_{\mathrm{s}2}\, \overline{v}_{\mathrm{s}3} u_{\mathrm{u}4}+ 2\, \overline{u}_{\mathrm{u}1}\gamma_5 v_{\mathrm{s}2}\, \overline{v}_{\mathrm{s}3}\gamma_5 u_{\mathrm{u}4} \big\}\nonumber\\
&\quad \hspace{4.5cm}\times b_{\mathrm{u}1}^\dagger d_{\mathrm{s}2}^\dagger d_{\mathrm{s}3} b_{\mathrm{u}4}.\label{eq:H_NJL_K_matrix}
\end{align}
Again, the number subscripts distinguish different spins and momenta while the summation indices are only the spin labels. 

We then evaluate the matrix elements of the NJL effective Hamiltonian given by Eq.~\eqref{eq:H_NJL_K_matrix} for the $K^+$ meson wavefunction. Explicitly, we have 
\begin{align}
&\quad  \big\langle\Psi'_{\mathrm{meson}+}(P'^+,\overrightarrow{P}'^\perp)\big\vert H_{\mathrm{NJL,K}}^{\mathrm{eff}} \big\vert\Psi_{\mathrm{meson}+}(P^+,\overrightarrow{P}^\perp)\big\rangle\nonumber\\
& =4\pi P^+ \delta(P'^+-P^+)(2\pi)^2\delta(\overrightarrow{P}'^\perp-\overrightarrow{P}^\perp)\nonumber\\
&\quad\times \sum_{s1's2's1s2}\int_{0}^{1}\dfrac{dx'}{4\pi x'(1-x')}\int_{0}^{1}\dfrac{dx}{4\pi x(1-x)}\nonumber\\
&\quad\times \int \dfrac{d\overrightarrow{\kappa}'^\perp}{(2\pi)^2}\int \dfrac{d\overrightarrow{\kappa}^\perp}{(2\pi)^2}\ \psi'^{\,*}_{s1's2'}(x',\kappa'^\perp)\psi_{s1s2}(x,\kappa^\perp)\,\nonumber\\
& \quad \times G_K\,\big\{- 2\,\overline{u}_{\mathrm{u}s1'}(p_1') v_{\mathrm{s}s2'}(p_2')\,\overline{v}_{\mathrm{s}s2}(p_2) u_{\mathrm{u}s1}(p_1) \nonumber\\
&\quad\quad + 2\,\overline{u}_{\mathrm{u}s1'}(p_1')\gamma_5 v_{\mathrm{s}s2'}(p_2')\,\overline{v}_{\mathrm{s}s2}(p_2)\gamma_5 u_{\mathrm{u}s1}(p_1) \big\},\label{eq:H_NJL_K_matrix_wf}
\end{align}
where the definitions of relative momenta are still given in Eq.~\eqref{eq:def_relative_momenta}. 

Similar to the case of the $\pi^+$-$\rho^+$ systems, within the basis representation defined by Eq.~\eqref{eq:psi_rs_basis_expansions}, we need to evaluate the following matrix elements ${\big\langle n'm'l's_1's_2'\big\vert \overline{u}_{\mathrm{u}} v_{\mathrm{s}}\,\overline{v}_{\mathrm{s}} u_{\mathrm{u}}\big\vert nmls_1s_2\big\rangle}$ and
${\big\langle n'm'l's_1's_2'\big\vert \overline{u}_{\mathrm{u}}\gamma_5 v_{\mathrm{s}}\,\overline{v}_{\mathrm{s}}\gamma_5 u_{\mathrm{u}}\big\vert nmls_1s_2\big\rangle}$. The explicit expressions of these matrix elements are also given in App.~\ref{ss:NJL_matrix_elements}.
\section{Results with basis truncations\label{sc:results}}
\subsection{Basis truncations and model parameters}
With the matrix elements of the effective Hamiltonian known exactly in our basis representation, the mass spectrum and the wavefunctions are to be solved numerically by diagonalizing the effective Hamiltonian. We truncate the basis representation by imposing maximums on the allowed basis modes. Specifically, the infinite summations in Eq.~\eqref{eq:psi_rs_basis_expansions} are replaced by the following finite sums:
\begin{equation}
\sum_{nml}\rightarrow \sum_{n=0}^{N_{\mathrm{max}}}\sum_{m=-M_{\mathrm{max}}}^{M_{\mathrm{max}}}\sum_{l=0}^{L_{\mathrm{max}}},\label{eq:basis_cutoffs}
\end{equation}
where $N_{\mathrm{max}}$, $M_{\mathrm{max}}$ and $L_{\mathrm{max}}$ are natural numbers specifying basis cutoffs. The truncations of the basis function given by Eq.~\eqref{eq:basis_cutoffs} indicate the existence of both infrared and ultraviolet regulators of our model. Because the off-diagonal matrix elements given by Tabs.~(\ref{tab:Heff_scalar_uuvv},~\ref{tab:Heff_scalar_uvvu},~\ref{tab:Heff_gamma5_uvvu}) do not couple to angular excitations with $\vert m\vert \geq 3$, we choose $M_{\mathrm{max}}=2$ as one natural basis cutoff on the orbital angular momentum. Meanwhile, we choose $N_{\mathrm{max}}$ as the cutoff on the transverse momentum. Doing so fixes the scale of our model. The default choice of the longitudinal cutoff is $L_{\mathrm{max}}=8$, except when calculating the meson PDAs. 

Roughly speaking, the infrared cutoff of our theory is $\Lambda_{\mathrm{IR}}=b/\sqrt{2N_{\mathrm{max}}+1}$, while the ultraviolet cutoff is $\Lambda_{\mathrm{UV}}=\sqrt{2N_{\mathrm{max}}+1}b$, with $b$ being the scale parameter in Eq.~\eqref{eq:def_phi_nm}~\cite{Li:2015zda,Li:2017mlw}. Because the low-energy effective descriptions of QCD are expected to change dramatically with model scales, in this article we do not explore extensively the dependence of our results on the transverse basis cutoffs. Also because our model is a low energy effective description, we expect it to be only valid for the ground states of the light mesons and we therefore restrict our application to these states. 

Within our modeling of the light meson systems, our calculated results show that the masses of the vector mesons $\rho^+$ and $K^{*+}$ are not sensitive to the coupling constants of the NJL interactions in Eqs.~(\ref{eq:H_eff_NJL_pi},~\ref{eq:H_eff_NJL_SU_3_ori}). Furthermore, the valence wavefunctions for the lowest vector states are almost entirely given by ${n=m=l=0}$ with spin triplet configurations:
\begin{align}
\psi_{rs}(x,\overrightarrow{\kappa}^\perp)& \simeq \chi_0(x) \,\phi_{00}\left(\dfrac{\overrightarrow{\kappa}^\perp}{\sqrt{x(1-x)}} \right)\nonumber\\
&\quad \times
\begin{cases}
\delta_{r+}\delta_{s+} \quad (\mathrm{for}\quad m_J=+1)\\
\dfrac{\delta_{r+}\delta_{s-}+\delta_{r-}\delta_{s+}}{\sqrt{2}}  \quad (\mathrm{for}\,m_J=0)\\
\delta_{r-}\delta_{s-} \quad (\mathrm{for}\quad m_J=-1)
\end{cases},\label{eq:wf_vector_meson_dominant}
\end{align}
where the total angular momentum projection is defined by $m_J=m+r+s$. The masses of the lightest vector meson with different $m_J$ are almost degenerate, and can be well approximated by the corresponding diagonal matrix elements in Eq.~\eqref{eq:diagonal_matrix_elements}, whereas the masses of the lowest pseudoscalar states are most sensitive to the coupling constants $G_{\pi}$ and $G_{K}$ in Eqs.~(\ref{eq:H_eff_NJL_pi},~\ref{eq:H_eff_NJL_SU_3_ori}). 

We work in the limit of the $\mathrm{SU}(2)$ flavor symmetry where the up quark mass and the down quark mass are identical, while the strange quark is expected to be heavier than the light quarks. Because we do not have $\mathrm{SU}(3)$ flavor symmetry, scales of the $\pi$-$\rho$ system and those of the $K$-$K^*$ system are expected to be different. We therefore assign different confining strength $\kappa$, quark mass and antiquark mass to these two systems. 

Specifically, we have three free parameters in our model for the $\pi^+$-$\rho^+$ system. They are the light quark mass $m_{\mathrm{l}}$, the light-light confining strength $\kappa_{\mathrm{ll}}$, and the pseudoscalar binding strength $G_{\pi}$, while for the $K^{+}$-$K^{*+}$ system, we have four free parameters. They are the light quark mass $m_{\mathrm{l}}$, the strange quark mass $m_{\mathrm{s}}$, the light-strange confining strength $\kappa_{\mathrm{ls}}$, and the pseudoscalar binding strength $G_{K}$.

In order to fix these parameters, we use the $\rho^+$ mass and the $K^{*+}$ mass as two constraints for the quark masses and system confining parameters (using an ``$\mathrm{l}$'' subscript to label a light quark) $m_{\mathrm{l}},~m_{\mathrm{s}},~\kappa_{\mathrm{ll}}$ and $\kappa_{\mathrm{ls}}$. Specifically since the vector states are almost entirely given by the ${n=m=l=0}$ states, we have from Eq.~\eqref{eq:H0_def} that for the $\pi^+$-$\rho^+$ system
\begin{equation}
4m_{\mathrm{l}}^2+3\kappa^2_{\mathrm{ll}}\simeq m_{\rho}^2,\label{eq:constraint_M_rho}
\end{equation}
while the corresponding relation in the $K^+$-$K^{*+}$ system is given by
\begin{equation}
\begin{cases}
4m_{\mathrm{l}}^2+3\kappa^2_{\mathrm{ls}}\simeq m_{\rho}^2\\
(m_{\mathrm{l}}+m_{\mathrm{s}})^2+3\kappa_{\mathrm{ls}}^2\simeq m_{K*}^2
\end{cases},\label{eq:constraint_M_K*}
\end{equation}
with $m_{\mathrm{l}}$ in Eq.~(\ref{eq:constraint_M_K*}) allowed to be different from the one in Eq.~\eqref{eq:constraint_M_rho}. The mass of the $\pi^+$ then fixes $G_{\pi}$. While the mass of $K^+$ determines $G_{K}$. We then choose the confining strength such that the charge radii of the $\pi^+$ and $K^+$ agree with experiments. The difference in the sizes of the $\pi^+$ and the $K^+$ implies different cutoffs on the dressing of the light quarks. Such dressing effects are approximated by Eqs.~(\ref{eq:constraint_M_rho},~\ref{eq:constraint_M_K*}) to reproduce the $\rho^+$ mass. At different confining strengths, this effective treatment accounts for the constituent light quarks being heavier in the $\pi^+$ and the $\rho^+$ than in the $K^+$ and the $K^{*+}$. 

The resulting model parameters for the $\pi^+$-$\rho^+$ system are given in Tab.~\ref{tab:input_pirho}. Our model reproduces the experimental $\pi^+$ mass and the $\rho^+$ mass as shown in Tab.~\ref{tab:static_perperties_pi_rho}. The uncertainty in the $\rho^+$ mass of our model comes from the slight splitting of the $\rho^+$ states with different angular momentum projections, while the model parameters for the $K^{+}$-$K^{*+}$ system are given in Tab.~\ref{tab:input_Kaons}. Similarly, we reproduce the $K^+$ mass, with the uncertainty in the $K^{*+}$ mass due to sensitivity to the total angular momentum projections. Our model for the $K^+$ mesons also finds a scalar state with the mass of $858.35~\mathrm{MeV}$. See Tab.~\ref{tab:static_perperties_K_K*} for details.
\begin{table}
	\centering
	\begin{tabular}{cccccc}
		\hline\hline
		$m_{\mathrm{l}}$ & $\kappa_{\mathrm{ll}}$ & $G_{\pi}$ & $N_{\mathrm{max}}$ & $M_{\mathrm{max}}$ & $L_{\mathrm{max}}$ \\ 
		\hline 
		$337.01~\mathrm{MeV}$ & $227.00~\mathrm{MeV}$ & $18.5095~\mathrm{GeV}^{-2}$ & $8$ & $2$ & $8$ \\ 
		\hline \hline
	\end{tabular} 
	\caption{Input parameters for the $\pi^+$-$\rho^+$ system of the BLFQ-NJL model. The corresponding basis cutoff scales are $\Lambda_{\mathrm{IR}}=55.06\,\mathrm{MeV}$ and $\Lambda_{\mathrm{UV}}=935.9\,\mathrm{MeV}$.}\label{tab:input_pirho}
\end{table}
\begin{table}
	\centering
	\begin{tabular}{ccc}
		\hline \hline 
		Parameter & BLFQ-NJL & PDG \\
		\hline 
		$m_{\pi+}$ & $139.57~\mathrm{MeV}$ & $139.57~\mathrm{MeV}$ \\
		$m_{\rho+}$ & $775.23\pm 0.04~\mathrm{MeV}$ & $775.26\pm0.25~\mathrm{MeV}$ \\
		$f_{\pi}$ & $202.10~\mathrm{MeV}$ & $130.2\pm1.7~\mathrm{MeV}$ \\
		$f_{\rho}$ & $100.12~\mathrm{MeV}$ & $221\pm 2\,\mathrm{MeV}$ \\
		$\sqrt{\langle r_c^2 \rangle}\vert _{\pi+}$ & $0.68\pm0.05~\mathrm{fm}$ & $0.672\pm0.008~\mathrm{fm}$ \\
		\hline \hline
	\end{tabular}
	\caption{Mass spectrum, decay constants, and the charge radii of the $\pi^+$ and $\rho^+$ ground states. The BLFQ-NJL results are obtained using parameters in Tab.~\ref{tab:input_pirho}. The small uncertainty in the calculated vector meson mass is due to splitting among states with different total angular momentum projections. The uncertainty of the pseudoscalar state charge radius reflects the error in the numerical evaluation of Eq.~\eqref{eq:def_charge_radius_pseudoscalar}. The experimental decay constant for the $\rho$ meson is extracted by Refs.~\cite{Maris:1999nt,Bhagwat:2006pu}. Other PDG results are from Ref.~\cite{Tanabashi:2018oca}.}\label{tab:static_perperties_pi_rho}
\end{table}
\begin{table}
	\centering
	\begin{tabular}{cccc}
		\hline \hline
		$m_{\mathrm{l}}$ & $m_{\mathrm{s}}$ & $\kappa_{\mathrm{ls}}$ & $G_K$ \\ 
		\hline 
		$307.66~\mathrm{MeV}$ & $445.14~\mathrm{MeV}$ & $276.00~\mathrm{MeV}$ & $13.6455~\mathrm{GeV}^{-2}$ \\ 
		\hline \hline
		\multicolumn{2}{c}{$N_{\mathrm{max}}$} & $M_{\mathrm{max}}$ & $L_{\mathrm{max}}$  \\
		\hline 
		\multicolumn{2}{c}{$8$} & $2$ & $8$ \\
		\hline \hline
	\end{tabular} 
	\caption{Input parameters for the $K^+$-$K^{*+}$ system of the BLFQ-NJL model. The corresponding basis cutoff scales are $\Lambda_{\mathrm{IR}}=66.94\,\mathrm{MeV}$ and $\Lambda_{\mathrm{UV}}=1138\,\mathrm{MeV}$.}\label{tab:input_Kaons}
\end{table}
\begin{table}
	\centering
	\begin{tabular}{ccc}
		\hline \hline 
		Parameter & BLFQ-NJL & PDG \\
		\hline 
		$m_{K+}$ & $493.68~\mathrm{MeV}$ & $493.68\pm0.02~\mathrm{MeV}$ \\
		$m_{K*+}$ & $891.82\pm 0.06~\mathrm{MeV}$ & $891.76\pm0.25~\mathrm{MeV}$ \\
		$m_{K_0^{*+}}$ & $858.35~\mathrm{MeV}$ & $824\pm30~\mathrm{MeV}$ \\
		$f_{K}$ & $235.99~\mathrm{MeV}$ & $155.6\pm0.4~\mathrm{MeV}$ \\
		$f_{K*}$ & $104.57~\mathrm{MeV}$ & $224\pm 11~\mathrm{MeV}$ \\
		$\sqrt{\langle r_c^2 \rangle}\vert _{K+}$ & $0.54\pm0.03~\mathrm{fm}$ & $0.560\pm 0.031~\mathrm{fm}$ \\
		\hline \hline 
	\end{tabular}
	\caption{Mass spectrum, decay constants, and the charge radius of the $K^+$ and $K^{*+}$ ground states. The BLFQ-NJL results are obtained using parameters in Tab.~\ref{tab:input_Kaons}. The uncertainty of the vector meson mass is due to splitting among states with different total angular momentum projections. The uncertainty of the pseudoscalar state charge radius reflects the error in the numerical evaluation of Eq.~\eqref{eq:def_charge_radius_pseudoscalar}. The experimental decay constant for the $K^*$ meson is extracted by Refs.~\cite{Maris:1999nt,Bhagwat:2006pu}. Other PDG results are from Ref.~\cite{Tanabashi:2018oca}.}\label{tab:static_perperties_K_K*}
\end{table}
\subsection{Calculations of decay constants, elastic form factors, and parton distribution amplitudes}
\subsubsection{Elastic form factor and charge radius}
To calculate the elastic form factors from the light front wavefunctions within the impulse approximation, we apply the Drell-Yan-West formula~\cite{Drell:1969km,West:1970av} within the Drell-Yan frame $P'^{+}=P^+$:
\begin{align}
&\quad I_{m_J,m_{J'}}(Q^2)= \Big\langle \Psi (P',m_J')\Big\vert \frac{J^+(0)}{2P^+} \Big\vert \Psi(P,m_J) \Big\rangle\nonumber\\
& =\sum_{rs}\int \dfrac{dx}{4\pi x(1-x)}\int \dfrac{d^2 k^\perp}{(2\pi)^2}\nonumber\\
&\quad \times \bigg\{\mathrm{e}_{\overline{\mathrm{q}}}\,\psi^{*\,m_{J'}}_{rs}\left(x,\overrightarrow{k}^\perp+(1-x)\overrightarrow{q}^\perp\right) \nonumber\\
&\quad \hspace{0.5cm} +\mathrm{e}_{\mathrm{q}} \psi^{*\,m_{J'}}_{rs}\left(x,\overrightarrow{k}^\perp-x\overrightarrow{q}^\perp\right)\bigg\} \psi^{m_{J}}_{rs}\left(x,\overrightarrow{k}^\perp\right),\label{eq:def_I_mj,mj'}
\end{align}
with $q=P'-P$ and $Q^2=-q^2$. Here $e_{\overline{\mathrm{q}}}$ is the charge of the antiquark, while $e_{\mathrm{q}}$ is the charge of the quark. The elastic form factors of the pseudoscalar states are then given by 
\begin{equation}
F_{\mathrm{P}}(Q^2)=I_{0,0}(Q^2).\label{eq:def_F_pseudoscalar}
\end{equation}
The charge radius is then defined through the first Taylor expansion coefficient of the elastic form factor at the origin:
\begin{equation}
\langle r_c^2 \rangle= -6\lim\limits_{Q^2\rightarrow 0}\,\dfrac{d}{dQ^2}\,F_{\mathrm{P}}(Q^2).\label{eq:def_charge_radius_pseudoscalar}
\end{equation}

\begin{figure*}
	\centering
	\includegraphics[width=1\linewidth]{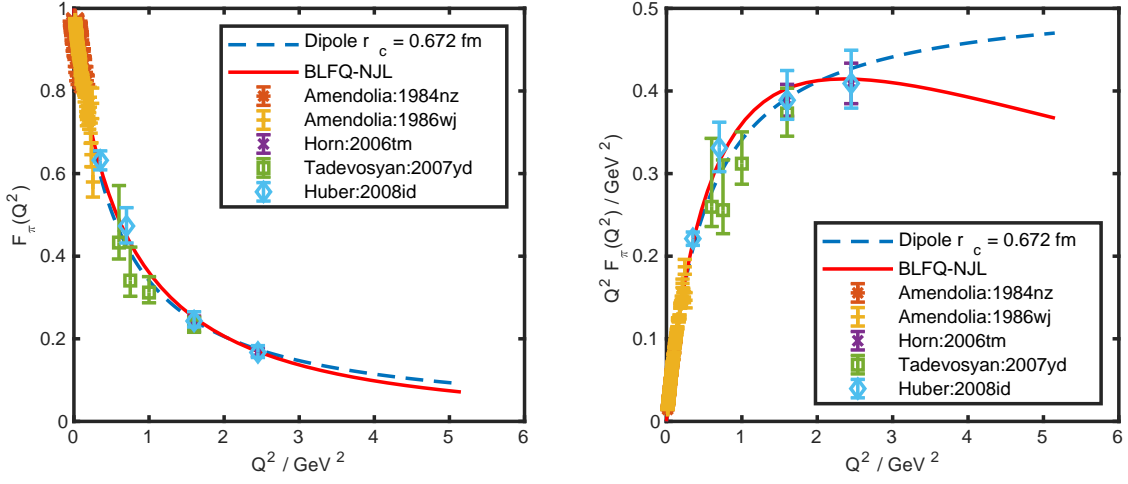}
	\caption{The elastic form factor for the $\pi^+$. The red solid line is the result from the BLFQ-NJL model. The blue dashed line is the dipole form with a charge radius of $0.672~\mathrm{fm}$. The orange stars, yellow plus signs, purple crosses, green boxes, and light blue diamonds with error bars correspond to experimental measurements from Refs.~\cite{Amendolia:1984nz,Amendolia:1986wj,Horn:2006tm,Tadevosyan:2007yd,Huber:2008id} respectively. The elastic form factor of the $\pi^+$ is plotted on the left panel. On the right panel, the same form factor is multiplied by $Q^2$.}
	\label{fig:F_pi}
\end{figure*}
\begin{figure*}
	\centering
	\includegraphics[width=1\linewidth]{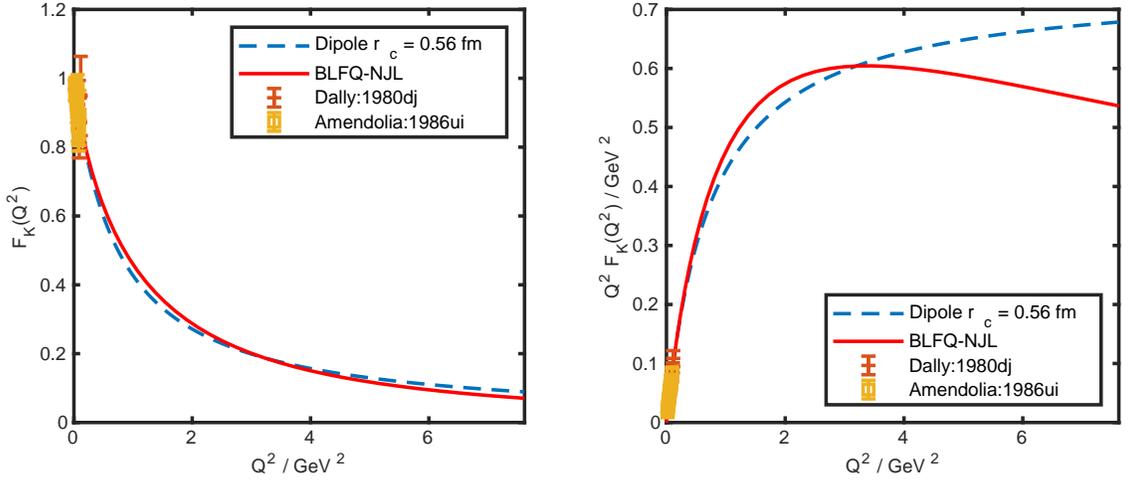}
	\caption{The elastic form factor for the $K^+$. The red solid line is the result from the BLFQ-NJL model. The blue dashed line is the dipole form with a charge radius of $0.560~\mathrm{fm}$. The orange plus signs and yellow boxes with error bars correspond to experimental measurements from Refs.~\cite{Dally:1980dj,Amendolia:1986ui} respectively.  The elastic form factor of the $K^+$ is plotted on the left panel. On the right panel, the same form factor is multiplied by $Q^2$.}
	\label{fig:F_K}
\end{figure*}
We then evaluate Eq.~\eqref{eq:def_I_mj,mj'} numerically using the pseudoscalar state wavefunctions obtained by our model with parameters given in Tabs.~\ref{tab:input_pirho},~\ref{tab:input_Kaons}. The resulting elastic form factors for the $\pi^+$ meson and for the $K^{+}$ meson are illustrated in Fig.~\ref{fig:F_pi} and Fig.~\ref{fig:F_K} respectively. We calculate the charge radii of the $\pi^+$ and $K^+$ mesons by fitting the behaviors of the elastic form factors as quadratic functions of $Q^2$ at small ${Q^2=(0,\,1,\,4,\,9,\,16,\,25,\,36,\,49,\,64)\,\kappa^2/16}$. The resulting radii are listed in Tabs.~\ref{tab:static_perperties_pi_rho},~\ref{tab:static_perperties_K_K*}, with the $95\%$ confidence intervals defining the fitting uncertainties. By using them as constraints in our parameter fitting, our results for these charge radii are in close agreement with the experiments. 

The functional form of the dipole elastic form factors in Figs.~\ref{fig:F_pi}~and~\ref{fig:F_K} is $F(Q^2)=\left(1+\langle r_c^2 \rangle\, Q^2/6 \right)^{-1}$. For both the $\pi^+$ and the $K^+$, the elastic form factors based on our model crosses the dipole result twice, once at small $Q^2\simeq \kappa^2$ and another time at around $2~\mathrm{GeV}^2$ for the $\pi^+$ and around $3~\mathrm{GeV}^2$ for the $K^+$. We expect deviations in the elastic form factors to occur when $Q^2>6/\langle r_c^2 \rangle$, because only with small $Q^2$ is the form factor fixed by the charge radius. For the $\pi^+$, both our result and the dipole form agree with available experimental data. Experimental results for $F_K(Q^2)$ with $Q^2\geq 1\,\mathrm{GeV}^2$ are available, but are not included in Fig.~\ref{fig:F_K} because of the large experimental uncertainties~\cite{Carmignotto:2018uqj}. 
\subsubsection{Decay constant and parton distribution amplitude} We take the definition of the meson decay constants as the matrix elements of current operators between the vacuum and the meson wavefunctions. Specifically, the decay constants for scalar mesons, pseudoscalar mesons, vector mesons, and axial vector mesons are define by
\begin{subequations}
	\label{eq:decay_constants}
	\begin{align}
	& \langle 0 \vert \overline{\psi}\,\gamma^\mu\,\psi \vert \mathrm{S}(p)\rangle =p^\mu f_{\mathrm{S}},\\
	& \langle 0 \vert \overline{\psi}\,\gamma^\mu\gamma_5\,\psi \vert \mathrm{P}(p)\rangle =i\,p^\mu f_{\mathrm{P}},
	\end{align}
	\begin{align}
	& \langle 0 \vert \overline{\psi}\,\gamma^\mu\,\psi \vert \mathrm{V}(p)\rangle =\epsilon^{\mu}_{\lambda}(p)\,m_{\mathrm{V}} f_{\mathrm{V}},\\
	& \langle 0 \vert \overline{\psi}\,\gamma^\mu\gamma_5\,\psi \vert \mathrm{A}(p)\rangle =\epsilon^{\mu}_{\lambda}(p)\,m_{\mathrm{A}} f_{\mathrm{A}},
	\end{align}
\end{subequations}
respectively. Here the polarization vector for the vector and axial-vector mesons is defined as
\begin{equation}
\epsilon^{\mu}_{\lambda}(p)=
\begin{cases}
\left(\frac{p^+}{m_{\mathrm{V},\mathrm{A}}},\,\frac{\overrightarrow{p}^{\perp 2}-m_{\mathrm{V},\mathrm{A}}^2}{m_{\mathrm{V},\mathrm{A}}\,p^+},\,\dfrac{\overrightarrow{p}^{\perp} }{m_{\mathrm{V},\mathrm{A}}}\right)\quad \mathrm{for}\quad \lambda=0\\
\left(0,\,\dfrac{2\overrightarrow{e}^{\perp}_\lambda\cdot \overrightarrow{p}^\perp }{p^+},\, \overrightarrow{e}^{\perp}_{\lambda} \right) \quad \mathrm{for}\quad \lambda=\pm 1
\end{cases},
\end{equation} 
with $\overrightarrow{e}^{\perp}_{\pm}=(1,\pm i)/\sqrt{2}$.

In terms of the valence sector light front wavefunctions, the decay constants are then given by~\cite{Lepage:1980fj,Li:2017mlw}
\begin{subequations}
	\label{eq:decay_constant_valence_WF}
	\begin{align}
	f_{\mathrm{P},\mathrm{A}} & =2\sqrt{2N_{\mathrm{c}}}\int_{0}^{1}\dfrac{dx}{4\pi\sqrt{x(1-x)}}\int \dfrac{d^2 \kappa^\perp}{(2\pi)^2}\nonumber\\
	&\quad \times \left[\psi_{+-}\left(x,\overrightarrow{\kappa} ^\perp\right)-\psi_{-+}\left(x,\overrightarrow{\kappa} ^\perp\right) \right]\bigg\vert _{m_J=0},\label{eq:decay_constant_valence_WF_PA}\\
	f_{\mathrm{S},\mathrm{V}} & =2\sqrt{2N_{\mathrm{c}}}\int_{0}^{1}\dfrac{dx}{4\pi\sqrt{x(1-x)}}\int \dfrac{d^2 \kappa^\perp}{(2\pi)^2}\nonumber\\
	&\quad \times\left[\psi_{+-}\left(x,\overrightarrow{\kappa} ^\perp\right)+\psi_{-+}\left(x,\overrightarrow{\kappa} ^\perp\right) \right]\bigg\vert _{m_J=0},
	\end{align}
\end{subequations}
with the condition $m_J=m+s_1+s_2=0$ specifying that only the states with zero angular momentum projections are used in the calculation. Based on our model, the decay constants for the $\pi^+$ and $\rho^+$ are given in Tab.~\ref{tab:static_perperties_pi_rho}. The decay constants for the $K^+$ and $K^{*+}$ are given in Tab.~\ref{tab:static_perperties_K_K*}. We note that the decay constants calculated from our model are dependent upon the transverse basis cutoff. Specifically with ${N_{\mathrm{max}}=6}$, we have ${f_\pi=187.93\,\mathrm{MeV}}$ and ${f_K=220.36\,\mathrm{MeV}}$. Meanwhile with ${N_{\mathrm{max}}=10}$, we find ${f_\pi=213.35\,\mathrm{MeV}}$ and ${f_K=248.38\,\mathrm{MeV}}$. On the other hand, the decay constants for the vector states are insensitive to $N_{\mathrm{max}}$ due to the wavefunctions being dominated by the lowest basis state. The increase of the pseudoscalar ground state decay constants with an increase in $N_{\mathrm{max}}$ is also observed for heavy mesons in Refs.~\cite{Li:2015zda,Li:2017mlw,Li:2018uif,Tang:2018myz}. Without an explanation of this behavior, we do not assign importance to our results on the decay constants. However, the ratio of the $K^+$ and $\pi^+$ decay constants from our model is ${f_K/f_{\pi}=1.168(4)}$, which is close to the experimental value ${f_K/f_{\pi}=1.1928(26)}$~\cite{Tanabashi:2018oca}. 

\begin{figure*}
	\centering
	\includegraphics[width=\linewidth]{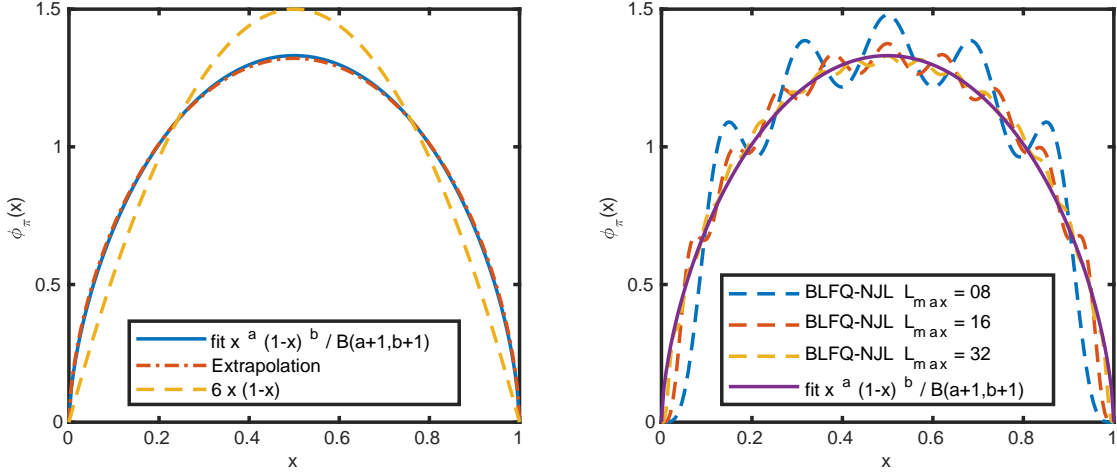}
	\caption{Parton distribution amplitude for the $\pi^+$ at the scale $\mu=935.9\,\mathrm{MeV}$. On the left panel, the blue solid line is the fit to the BLFQ-NJL model using Eq.~\eqref{eq:PDA_fits} with $a=b=0.62$. The red dot-dashed line corresponds to the same functional form with the $a$ and $b$ parameters extrapolated to $L_{\mathrm{max}}\rightarrow+\infty$. The orange dashed line represents the $6\,x(1-x)$ from perturbative QCD. On the right panel, PDAs for the $\pi^+$ with $L_{\mathrm{max}}=8,\,16,\,32$ are plotted in dashed line, together with the fit to the $L_{\mathrm{max}}=32$ case.}
	\label{fig:phi_pi}
\end{figure*}
\begin{figure*}
	\centering
	\includegraphics[width=\linewidth]{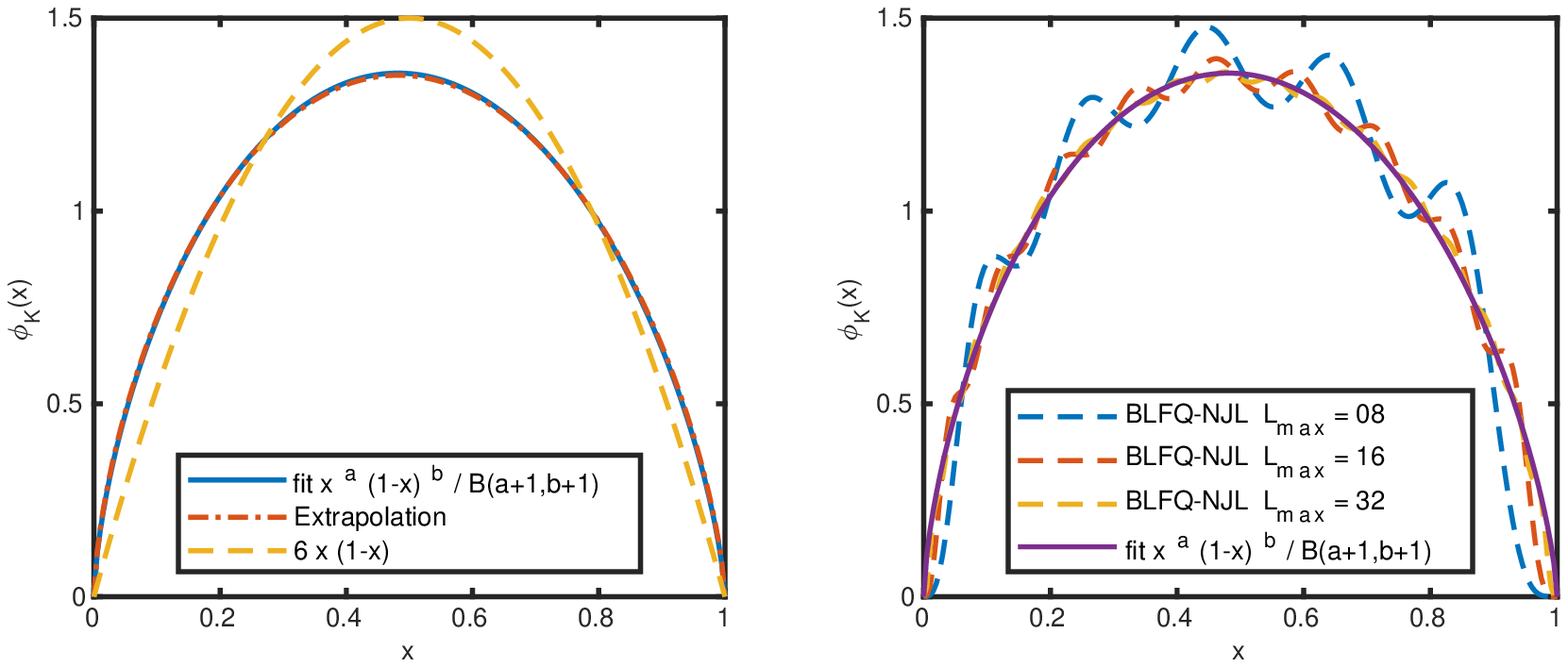}
	\caption{Parton distribution amplitude for the $K^+$ at the scale $\mu=1138\,\mathrm{MeV}$. On the left panel, the blue solid line is the fit to the BLFQ-NJL model using Eq.~\eqref{eq:PDA_fits} with $a=0.65$ and $b=0.70$. The red dot-dashed line corresponds to the same functional form with the $a$ and $b$ parameters extrapolated to $L_{\mathrm{max}}\rightarrow+\infty$. The red dashed line represents the $6\,x(1-x)$ from perturbative QCD in the $\mathrm{SU}(3)$ flavor symmetric limit. On the right panel, PDAs for the $K^+$ with $L_{\mathrm{max}}=8,\,16,\,32$ are plotted, together with the fit to the $L_{\mathrm{max}}=32$ case.}
	\label{fig:phi_K}
\end{figure*}
\begin{table*}
	\centering
	\begin{tabular}{ccccccccc}
		\hline \hline
		$L_{\mathrm{max}}$ & $8$ & $12$ & $16$ & $20$ & $24$ & $28$ & $32$ & Extrapolated to $+\infty$ \\ 
		\hline 
		$\pi^+~a=b$ & $0.7897$ & $0.6951$ & $0.6562$ & $0.6380$ & $0.6285$ & $0.6234$ & $0.6204$ & $0.5996$ \\ 
		$f_\pi~(\mathrm{MeV})$ & $202.10$ & $204.09$ & $204.92$ & $205.32$ & $205.53$ & $205.65$ & $205.72$ &  \\ 
		$K^+~a$ & $0.7589$ & $0.6969$ & $0.6727$ & $0.6609$ & $0.6575$ & $0.6513$ & $0.6507$ & $0.6398$ \\ 
		$K^+~b$ & $0.8413$ & $0.7594$ & $0.7282$ & $0.7130$ & $0.7089$ & $0.7006$ & $0.7001$ & $0.6874$ \\ 
		$f_K~(\mathrm{MeV})$ & $235.98$ & $238.01$ & $238.81$ & $239.20$ & $239.37$ & $239.51$ & $239.56$ &  \\ 
		\hline \hline
	\end{tabular} 
	\caption{Dependence of the PDA fitting parameters and the decay constants of $\pi^+$ and $K^+$ on the longitudinal basis cutoff $L_{\mathrm{max}}$. The extrapolations are carried out by fitting to quadratic functions of $L_{\mathrm{max}}^{-1}$}\label{tab:L_max_dep}
\end{table*}
\begin{figure*}
	\centering
	\includegraphics[width=\linewidth]{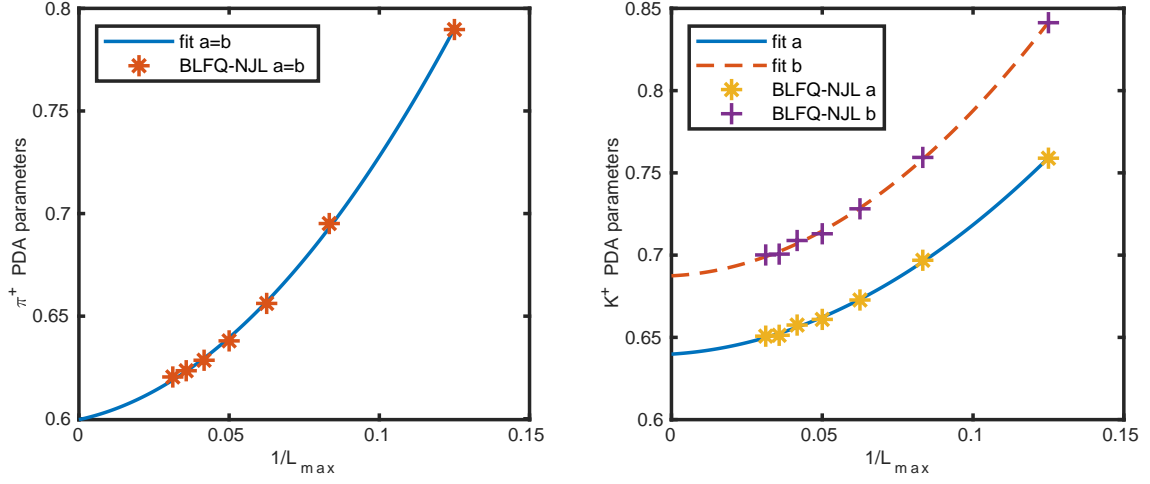}
	\caption{Extrapolations of the dependence of fitting parameters on the longitudinal basis cutoff using quadratic functions of $L_{\mathrm{max}}^{-1}$. On the left panel, the red stars are the parameters $a=b$ from fitting the BLFQ-NJL PDA for the $\pi^{+}$ using the functional form given by Eq.~\eqref{eq:PDA_fits}. The blue solid line is the fitted quadratic function of $L_{\mathrm{max}}^{-1}$. The right panel shows the extrapolation of fitting parameters for the $K^{+}$ PDA obtained using the BLFQ-NJL model. The orange stars are the parameters $a$ from fitting BLFQ-NJL using the functional form given by Eq.~\eqref{eq:PDA_fits}. The blue solid line is the fitted quadratic function of $L_{\mathrm{max}}^{-1}$ for the parameter $a$. The purple stars are the parameters $b$ from fitting the BLFQ-NJL PDA using the functional form given by Eq.~\eqref{eq:PDA_fits}. The red dashed line is the fitted quadratic function of $L_{\mathrm{max}}^{-1}$ for the parameter $b$.}
	\label{fig:phipikext}
\end{figure*}
The parton distribution amplitudes (PDAs) of the pseudoscalar states are defined as~\cite{Li:2017mlw} 
\begin{align}
\phi_{\mathrm{P}}(x;\mu) & =\dfrac{2\sqrt{2N_{\mathrm{c}}}}{f_{\mathrm{P}}}\dfrac{1}{4\pi\sqrt{x(1-x)}}\int \dfrac{d^2 \kappa^\perp}{(2\pi)^2}\nonumber\\
&\quad \times\left[\psi_{+-}\left(x,\overrightarrow{\kappa} ^\perp\right)-\psi_{-+}\left(x,\overrightarrow{\kappa} ^\perp\right) \right],\label{eq:def_PDA_P}
\end{align}
with the $f_{\mathrm{P}}$ being the decay constant defined by Eq.~\eqref{eq:decay_constant_valence_WF_PA} such that the PDAs are normalized to $1$. The PDA for the vector state is defined using the $m_J=0$ state as
\begin{align}
\phi_{\mathrm{V}}(x;\mu) & =\dfrac{2\sqrt{2N_{\mathrm{c}}}}{f_{\mathrm{V}}}\dfrac{1}{4\pi\sqrt{x(1-x)}}\int \dfrac{d^2 \kappa^\perp}{(2\pi)^2}\nonumber\\
&\quad \times \left[\psi_{+-}\left(x,\overrightarrow{\kappa} ^\perp\right)+\psi_{-+}\left(x,\overrightarrow{\kappa} ^\perp\right) \right]\bigg\vert_{m_J=0}.\label{eq:def_PDA_V}
\end{align}
Within the BLFQ-NJL model, the number of colors we use is $N_c=3$. Our basis truncation provides the ultraviolet cutoff of ${\mu= \sqrt{2N_{\mathrm{max}}+1} b}$ for the transverse integrals in Eqs.~(\ref{eq:def_PDA_P},~\ref{eq:def_PDA_V})~\cite{Li:2017mlw}. 

At the corresponding model scales, the PDAs for the $\pi^+$ and $K^+$ are given by Fig.~\ref{fig:phi_pi} and Fig.~\ref{fig:phi_K} respectively. Both PDAs contain multiple humps, the number of which is dependent on the longitudinal cutoff parameter $L_{\mathrm{max}}$. The pion PDA is symmetric about $x=0.5$, while the kaon PDA is skewed toward $m_{\mathrm{l}}/(m_{\mathrm{l}}+m_{\mathrm{s}})=0.4$. The results with different $L_{\mathrm{max}}$ are obtained with the same set of quark mass parameters, confining strength $\kappa$, $N_{\mathrm{max}}$ and $M_{\mathrm{max}}$ specified in Tabs.~\ref{tab:input_pirho},~\ref{tab:input_Kaons}. The NJL coupling constants $G_{\pi}$ and $G_{K}$ are adjusted such that the masses of the pseudoscalar and vector states are within $1\,\%$ of their experimental values. Notice that the scale $\mu$ of the PDAs is independent from the choices of $L_{\mathrm{max}}$, since the transverse cutoff is kept fixed at ${N_{\mathrm{max}}=8}$ . The scale for the $\pi^+$-$\rho^+$ system is $\mu=935.9\,\mathrm{MeV}$, while the scale for the $K^+$-$K^{*+}$ is $\mu=1138\,\mathrm{MeV}$. Therefore, humps on the right panels of Figs.~\ref{fig:phi_pi}~and~\ref{fig:phi_K} are numerical artifacts. 
\begin{figure*}
	\centering
	\includegraphics[width=\linewidth]{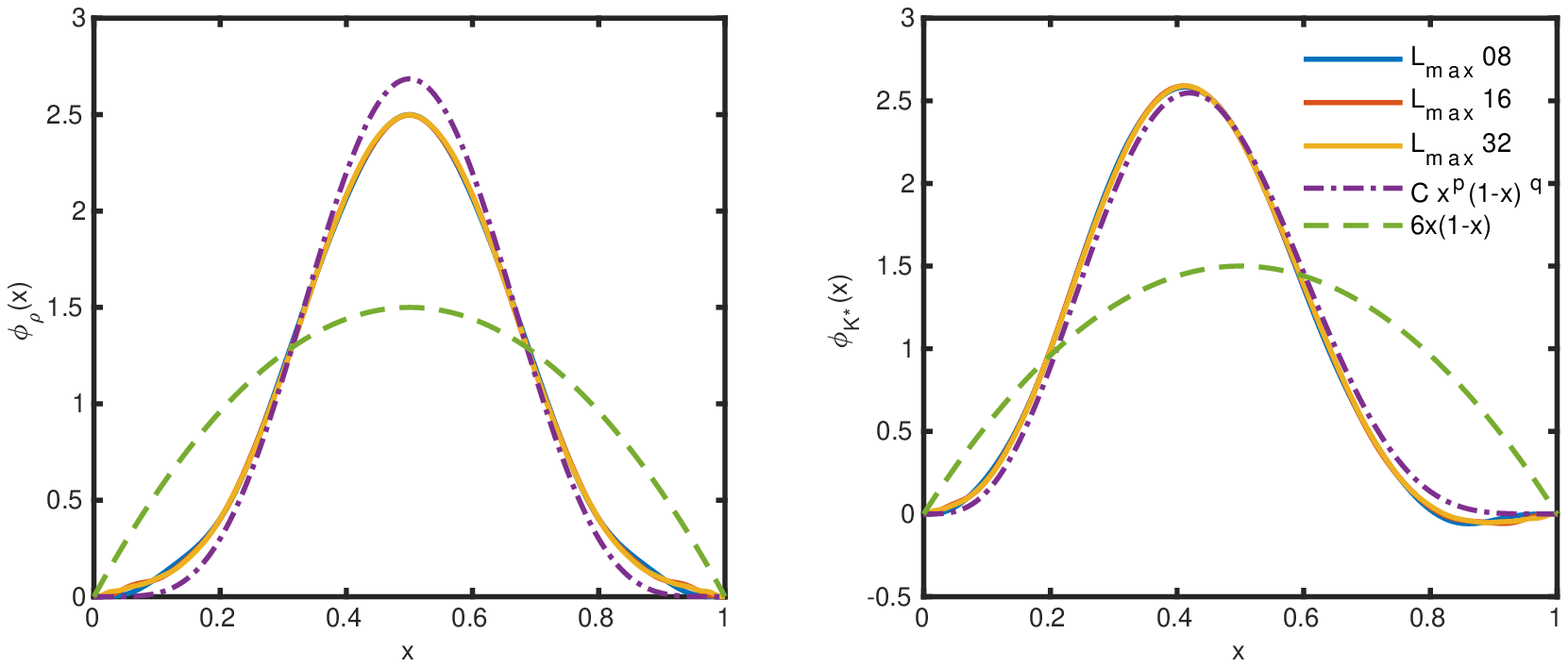}
	\caption{Left panel: parton distribution amplitude for the $\rho^+$ at the scale $\mu=935.9\,\mathrm{MeV}$. Right panel: parton distribution amplitude for the $K^{*+}$ at the scale $\mu=1138\,\mathrm{MeV}$. Legends are identical on both panels and therefore are only shown on the right. The green dashed line corresponds to $6\,x(1-x)$ from perturbative QCD in the $\mathrm{SU}(3)$ flavor symmetric limit. The blue, red and orange solid lines are PDAs obtained at $L_{\mathrm{max}}=8,\,16,\,32$ respectively. The purple dot-dashed line corresponds to the functional form in Eq.~\eqref{eq:PDA_fits} with parameters $p=(\beta+1)/2=4.91,\,q=(\alpha+1)/2=4.91$ for the $\rho^+$, and $p=(\beta+1)/2=3.54,\,q=(\alpha+1)/2=4.90$ for the $K^{*+}$. These $\alpha$ and $\beta$ are calculated by Eq.~\eqref{eq:def_alpha_beta} with parameters in Tabs.~\ref{tab:input_pirho},~\ref{tab:input_Kaons} to indicate the result with only the leading basis state contributing to the PDAs.}
	\label{fig:phi_rho_Ks}
\end{figure*}

Based on Figs.~\ref{fig:phi_pi},~\ref{fig:phi_K}, one observes that with an increase in $L_{\mathrm{max}}$, the PDAs for the pseudoscalar states trend toward a smooth function with decreasing oscillation amplitude about a single-peaked function. 
We therefore fit the PDAs for the $\pi^+$ and $K^+$ from our model with different $L_{\mathrm{max}}$ using the following functional form~\cite{Segovia:2013eca}
\begin{equation}
\phi(x)=\dfrac{x^a (1-x)^b}{B(a+1,b+1)}, \label{eq:PDA_fits}
\end{equation}
where $B(a+1,b+1)=\Gamma(a+1)\Gamma(b+1)/\Gamma(a+b+2)$ is the Euler Beta function. With $L_{\mathrm{max}}=32$, we find out that for the pion PDA the best fitting parameters are ${a=b=0.62}$. While for the kaon PDA with the same $L_{\mathrm{max}}$, the best fit is specified by ${a=0.65}$ and ${b=0.70}$. Fits are also performed for other values of $L_{\mathrm{max}}\in \{8,\,12,\,16,\,20,\,24,\,28,\,32\}$ and extrapolated to $L_{\mathrm{max}}\rightarrow +\infty$ by fitting the resulting $a$ and $b$ as quadratic functions of $L_{\mathrm{max}}^{-1}$. We find out the extrapolated values for the $\pi^+$ PDA are $a=b=0.60$, while for the $K^{+}$ PDA, the extrapolated values are $a=0.64$ and $b=0.69$. See Tab.~\ref{tab:L_max_dep} and Fig.~\ref{fig:phipikext} for details of these extrapolations. Then using these extrapolated values, the corresponding PDAs are given as the dot-dashed lines on the left panels of Figs.~\ref{fig:phi_pi}~and~\ref{fig:phi_K}, which are barely distinguishable from the $L_{\mathrm{max}}=32$ fitting results. We notice that the decay constants, the charge radii, and the elastic form factors each depend weakly on $L_{\mathrm{max}}$ as shown in Tab.~\ref{tab:L_max_dep}. 

We also present the PDAs for the $\rho^+$ and $K^{*+}$ vector mesons from the BLFQ-NJL model in Fig.~\ref{fig:phi_rho_Ks}. The convergence of the vector meson PDAs with respect to $L_{\mathrm{max}}$ is much faster than the case with the pseudoscaler mesons. In Fig.~\ref{fig:phi_rho_Ks}, we also show the PDAs calculated with only the dominant basis component specified by Eq.~\eqref{eq:wf_vector_meson_dominant}. The sub-dominant basis components of the $\rho^+$ light front wavefunction makes the PDA broader by the enhancements near $x=0$ and $x=1$. For the $K^{*+}$, the enhancement is also observed near $x=0$, with the additional feature that the PDA becomes slightly negative near $x=1$. 
\section{Summary\label{sc:summary}}
We have combined the chiral dynamics in the form of the Nambu--Jona-Lasinio model with the basis light front quantization framework. With our basis representation of the valence Fock sector light front wavefunction for the mesons, we calculated the matrix elements of the NJL interactions analytically. We then solved the ground state light front wavefunctions for the valence quarks of $\pi^+$, $\rho^+$, $K^+$, and $K^{*+}$ by diagonalizing the effective light front Hamiltonian with confinement potentials and the NJL interactions. The parameters of our model were adjusted to reproduce the experimental mass spectrum of these mesons. 

We found that the vector ground states were almost entirely given by one momentum space basis with spin triplet configurations, while the pseudoscalar states were complex mixtures of many basis functions. We calculated decay constants for $\pi^+$, $\rho^+$, $K^+$ and $K^{*+}$, and found that the ratio $f_{\pi}/f_{K}$ was close to the experimental value. Specifically for the $\pi^+$ and $K^{+}$, we presented our results for the elastic form factors along with comparison to available experimental data. We calculated the charge radii of these two states by analyzing the small $Q^2$ behaviors of the corresponding elastic form factors. The choice of the scale parameters of our model were optimized through adjusting the charge radii of the pseudoscalar states to agree with experimental data.

We also calculated the parton distribution amplitudes for $\pi^+$, $\rho^+$, $K^{+}$ and $K^{*+}$ at our model scales. We illustrated their good convergence on the longitudinal cutoff parameter $L_{\mathrm{max}}$. For the PDAs of the pseudoscalar states, we fitted their PDAs obtained at different $L_{\mathrm{max}}$ by the functional form in Eq.~\eqref{eq:PDA_fits}. We then extrapolated the results to $L_{\mathrm{max}}\rightarrow+\infty$. The PDAs for the $\rho^+$ and $K^{*+}$ vector states were also presented. They showed faster convergence with respect to $L_{\mathrm{max}}$, with differences to the results using the dominant light front basis functions illustrated. 

We expect to generalize our NJL-BLFQ model to describe the structures of flavor singlet light mesons, including $\pi^0$, $\eta$, $\eta'$, $\rho^0$, $\omega$, and $\phi$. The flavor neutral terms of the NJL interactions indicate nontrivial flavor wavefunctions for the valence quarks of these mesons. With flavor contents of the wavefunctions to be determined by the Hamiltonian, the mixing between $K^0$ and $\overline{K}^0$ is another interesting feature to explore. 
\section*{Acknowledgments}
We would like to thank Pieter Maris for suggestions leading to our inclusion of additional results. We also appreciate discussions with Xingbo Zhao, Wenyang Qian, Yang Li, Meijian Li, Shuo Tang, and Anji Yu on the general topics of BLFQ. This work was supported by the Department of Energy under Grants No. DE-FG02-87ER40371, No. DE-SC0018223 (SciDAC4/NUCLEI), and No. DE-SC0015376 (DOE Topical Collaboration in Nuclear Theory for Double-Beta Decay and Fundamental Symmetries).
\appendix
\section{Spin dependence of the NJL interactions in BLFQ\label{ss:NJL_matrix_elements}}
\subsection{Dirac spinors and bilinears\label{ss:spinor_def}}
We adopt the Weyl representation of the Dirac matrices. They are explicitly given by
\begin{equation}
\gamma^0=\begin{pmatrix}
\mathbf{0} & \mathbf{1} \\ 
\mathbf{1} & \mathbf{0}
\end{pmatrix}\quad \mathrm{and}\quad \gamma^i=\begin{pmatrix}
\mathbf{0} & \sigma^i \\ 
-\sigma^i & \mathbf{0}
\end{pmatrix},
\end{equation}
with $i\in \{1,~2,~3\}$. Here $\sigma^i$ are Pauli matrices defined as 
\begin{equation}
\sigma^1=\begin{pmatrix}
0 & 1 \\ 
1 & 0
\end{pmatrix} ,\quad 
\sigma^2=\begin{pmatrix}
0 & -i \\ 
i & 0
\end{pmatrix} ,\quad 
\sigma^1=\begin{pmatrix}
1 & 0 \\ 
0 & -1
\end{pmatrix} .
\end{equation}
One can then verify that
\begin{equation}
\sigma^i\sigma^j=\mathbf{1}\delta^{ij}+i\epsilon^{ijk}\sigma^k,
\end{equation}
where $\epsilon^{ijk}$ is the Levi-Civita symbol. Consequently, we have $\{\gamma^\mu,\gamma^\nu\}=2g^{\mu\nu}$, with $g^{00}=+1$ and $g^{ii}=-1$ for $i=1,~2,~3$. Additionally, the light front gamma matrices are defined as $\gamma^{\pm}=\gamma^0\pm \gamma^3$.

The Dirac spinors are then given by
\begin{subequations}
	\begin{align}
	u_{\pm 1/2}(p)&=\dfrac{1}{2\sqrt{p^+}}(\slashed{p}+m)\gamma^+\chi_{\pm 1/2}\\
	v_{\pm 1/2}(p)&=\dfrac{1}{2\sqrt{p^+}}(\slashed{p}-m)\gamma^+\chi_{\mp 1/2},
	\end{align}
\end{subequations}
with the spin vectors defined as
\begin{equation}
\chi_{+1/2}=(0,0,1,0)^{\mathrm{T}},\quad \chi_{-1/2}=(0,1,0,0)^{\mathrm{T}}.
\end{equation}
We can then calculate the spin decomposition of the Dirac bilinears. Specifically, we need to calculate the following three combinations: $\overline{u}_{s1'}(p_1')u_{s1}(p_1)\,\overline{v}_{s2}(p_2)v_{s2'}(p_2')$, $\overline{u}_{s1'}(p_1')v_{s2'}(p_2')\, \overline{v}_{s2}(p_2)u_{s1}(p_1)$, and $\overline{u}_{s1'}(p_1')\gamma_5 v_{s2'}(p_2')\, \overline{v}_{s2}(p_2)\gamma_5 u_{s1}(p_1)$. The explicit spin decompositions of these three terms are given in Tabs.~\ref{tab:spin_uuvv},~\ref{tab:spin_uvvu},~and~\ref{tab:spin_gamma5_qLR}. In these tables, $\mathbf{m}$ is the mass of the quark. While $\overline{\mathbf{m}}$ is the mass of the antiquark. The transverse momenta $\overrightarrow{q}^\perp$ and $\overrightarrow{q}'^\perp$ are defined by
\begin{equation}
\overrightarrow{\kappa}'^\perp=\sqrt{x'(1-x')}\overrightarrow{q}'^\perp\quad\mathrm{and}\quad \overrightarrow{\kappa}^\perp=\sqrt{x(1-x)}\overrightarrow{q}^\perp,
\end{equation}
with $\overrightarrow{\kappa}^\perp$, $\overrightarrow{\kappa}'^\perp$, $x$ and $x'$ given in Eq.~\eqref{eq:def_relative_momenta}. The left and right transverse momenta are defined as $q^{\mathrm{L}}=q_1-iq_2$ and $q^{\mathrm{R}}=q_1+iq_2$.
\bgroup
\def\arraystretch{1.5}
\begin{table}
	\centering
	\begin{tabular}{c|c}
		\hline \hline $s_1's_2's_2s_1$ & $\overline{u}_{\mathrm{u}s1'}(p_1')u_{\mathrm{u}s1}(p_1)\,\overline{v}_{\mathrm{d}s2}(p_2)v_{\mathrm{d}s2'}(p_2')$ \\
		\hline $++++$ & $-\mathbf{m}\mathbf{\overline{m}}\left(\sqrt{\dfrac{x'}{x}}+\sqrt{\dfrac{x}{x'}}\right) \left(\sqrt{\dfrac{1-x'}{1-x}}+\sqrt{\dfrac{1-x}{1-x'}} \right) $ \\
		\hline $+++-$ & $\mathbf{\overline{m}}\left(\sqrt{\dfrac{x'}{1-x'}}q^{\mathrm{L}}-\sqrt{\dfrac{x}{1-x}}q'^{\mathrm{L}} \right)(2-x'-x)$ \\ 
		\hline $++-+$ & $\mathbf{m}(x'+x)\left(\sqrt{\dfrac{1-x}{x}}q'^{\mathrm{L}}-\sqrt{\dfrac{1-x'}{x'}}q^{\mathrm{L}} \right)$ \\ 
		\hline $++--$ & $\begin{aligned} & -(x'+x-2x'x)q'^{\mathrm{L}}q^{\mathrm{L}} \\ &\quad +\sqrt{x'(1-x')x(1-x)}(q'^{\mathrm{L}2}+q^{\mathrm{L}2}) \end{aligned}$ \\ 
		\hline $+-++$ & $\mathbf{m}(x'+x)\left(\sqrt{\dfrac{1-x'}{x'}}q^{\mathrm{R}}-\sqrt{\dfrac{1-x}{x}}q'^{\mathrm{R}} \right)$ \\ 
		\hline $+-+-$ & $\begin{aligned} & (1-x')xq'^{\mathrm{L}}q^{\mathrm{R}}+x'(1-x)q'^{\mathrm{R}}q^{\mathrm{L}}\\ & \quad -\sqrt{x'(1-x')x(1-x)}(q'^{\mathrm{L}}q'^{\mathrm{R}}+q^{\mathrm{L}}q^{\mathrm{R}} )\end{aligned}$ \\ 
		\hline $+--+$ & $++++$ \\ 
		\hline $+---$ & $+++-$ \\ 
		\hline $-+++$ & $-\mathrm{\overline{\mathbf{m}}}\left(\sqrt{\dfrac{x'}{1-x'}}q^{\mathrm{R}}-\sqrt{\dfrac{x}{1-x}}q'^{\mathrm{R}} \right)(2-x'-x)$ \\ 
		\hline $-++-$ & $++++$ \\ 
		\hline $-+-+$ & $\begin{aligned}
		& 
		x'(1-x)q'^{\mathrm{L}}q^{\mathrm{R}}+x(1-x')q'^{\mathrm{R}}q^{\mathrm{L}}\\&\quad-\sqrt{x'(1-x')x(1-x)}(q'^{\mathrm{L}}q'^{\mathrm{R}}+q^{\mathrm{L}}q ^{\mathrm{R}}) \end{aligned}$ \\ 
		\hline $-+--$ & $++-+$ \\ 
		\hline $--++$ & $\begin{aligned} &-(x'+x-2x'x)q'^{\mathrm{R}}q^{\mathrm{R}}\\&\quad+\sqrt{x'(1-x')x(1-x)}(q'^{\mathrm{R}2}+q^{\mathrm{R}2}) \end{aligned}$ \\ 
		\hline $--+-$ & $+-++$ \\ 
		\hline $---+$ & $-+++$ \\ 
		\hline $----$ & $++++$ \\ 
		\hline \hline 
	\end{tabular} 
	\caption{The spin dependence of the scalar bilinear product $\overline{u}u\,\overline{v}v$ written in terms of momentum fractions and relative momenta.}
	\label{tab:spin_uuvv}
\end{table}
\egroup
\bgroup
\def\arraystretch{1.5}
\begin{table}
	\centering
	\begin{tabular}{c|c}
		\hline \hline $s_1's_2's_2s_1$ & $\overline{u}_{\mathrm{u}s1'}(p_1')v_{\mathrm{s}s2'}(p_2')\,\overline{v}_{\mathrm{s}s2}(p_2)u_{\mathrm{u}s1}(p_1)$ \\
		\hline $++++$ & $q'^{\mathrm{L}}q^{\mathrm{R}} $ \\
		\hline $+++-$ & $q'^{\mathrm{L}}\left(\sqrt{\dfrac{1-x}{x}}\mathbf{m}-\sqrt{\dfrac{x}{1-x}}\overline{\mathbf{m}} \right)$ \\ 
		\hline $++-+$  & $+++-$ \\ 
		\hline $++--$ & $ -q'^{\mathrm{L}}q^{\mathrm{L}}$ \\ 
		\hline $+-++$ & $\left(\sqrt{\dfrac{1-x'}{x'}}\mathbf{m}-\sqrt{\dfrac{x'}{1-x'}}\overline{\mathbf{m}} \right)q^{\mathrm{R}}$ \\ 
		\hline $+-+-$ & $\begin{aligned} & \left(\sqrt{\dfrac{1-x'}{x'}}\mathbf{m}-\sqrt{\dfrac{x'}{1-x'}}\overline{\mathbf{m}}\right)\\
		& \quad\times\left(\sqrt{\dfrac{1-x}{x}}\mathbf{m}-\sqrt{\dfrac{x}{1-x}}\overline{\mathbf{m}}\right) \end{aligned}$ \\ 
		\hline $+--+$ & $+-+-$ \\ 
		\hline $+---$ & $-\left( \sqrt{\dfrac{1-x'}{x'}}\mathbf{m}-\sqrt{\dfrac{x'}{1-x'}}\overline{\mathbf{m}} \right)q^{\mathrm{L}}$ \\ 
		\hline $-+++$ & $+-++$ \\ 
		\hline $-++-$ & $+-+-$ \\ 
		\hline $-+-+$ & $+-+-$ \\ 
		\hline $-+--$ & $+---$ \\ 
		\hline $--++$ & $-q'^{\mathrm{R}}q^{\mathrm{R}}$ \\ 
		\hline $--+-$ & $-q'^{\mathrm{R}}\left(\sqrt{\dfrac{1-x}{x}}\mathbf{m}-\sqrt{\dfrac{x}{1-x}}\overline{\mathbf{m}} \right)$ \\ 
		\hline $---+$ & $--+-$ \\ 
		\hline $----$ & $q'^{\mathrm{R}}q^{\mathrm{L}}$ \\ 
		\hline \hline 
	\end{tabular} 
	\caption{The spin dependence of the scalar bilinear product $\overline{u} v\,\overline{v}u$ written in terms of momentum fractions and relative momenta.}
	\label{tab:spin_uvvu}
\end{table}
\egroup
\bgroup
\def\arraystretch{1.5}
\begin{table}\centering
	\begin{tabular}{c|c}
		\hline \hline $s'_1s'_2s_2s_1$ & $\overline{u}_{s1'}(p_1')\gamma^5v_{s2'}(p_2')\,\overline{v}_{s2}(p_2)\gamma^5 u_{s1}(p_1)$ \\ 
		\hline $++++$ & $-q'^{\mathrm{L}} q^{\mathrm{R}}$ \\ 
		\hline $+++-$ & $-\left(\sqrt{\dfrac{1-x}{x}}\mathrm{\mathbf{m}}+\sqrt{\dfrac{x}{1-x}}\overline{\mathrm{\mathbf{m}}}\right)q'^{\mathrm{L}}$ \\ 
		\hline $++-+$ & $(-1)\,+++-$ \\ 
		\hline $++--$ & $-q'^{\mathrm{L}} q^{\mathrm{L}}$ \\ 
		\hline $+-++$ & $\left(\sqrt{\dfrac{1-x'}{x'}}\mathrm{\mathbf{m}}+\sqrt{\dfrac{x'}{1-x'}}\overline{\mathrm{\mathbf{m}}}\right)q^{\mathrm{R}}$ \\ 
		\hline $+-+-$ & $\dfrac{[x'\overline{\mathrm{\mathbf{m}}}+(1-x')\mathrm{\mathbf{m}}][x\overline{\mathrm{\mathbf{m}}}+(1-x)\mathrm{\mathbf{m}}]}{\sqrt{x'(1-x')x(1-x)}}$ \\ 
		\hline $+--+$ & $(-1)\,+-+-$ \\ 
		\hline $+---$ & $\left(\sqrt{\dfrac{x'}{1-x'}}\overline{\mathrm{\mathbf{m}}}+\sqrt{\dfrac{1-x'}{x'}}\mathrm{\mathbf{m}}\right)q^{\mathrm{L}}$ \\ 
		\hline $-+++$ & $(-1)\,+-++$ \\ 
		\hline $-++-$ & $(-1)\,+-+-$ \\ 
		\hline $-+-+$ & $+-+-$ \\ 
		\hline $-+--$ & $(-1)\,+---$ \\ 
		\hline $--++$ & $-q'^{\mathrm{R}} q^{\mathrm{R}}$ \\ 
		\hline $--+-$ & $-\left(\sqrt{\dfrac{x}{1-x}}\overline{\mathrm{\mathbf{m}}}+\sqrt{\dfrac{1-x}{x}}\mathrm{\mathbf{m}}\right)q'^{\mathrm{R}}$ \\ 
		\hline $---+$ & $(-1)\,--+-$ \\ 
		\hline $----$ & $-q'^{\mathrm{R}} q^{\mathrm{L}}$ \\ 
		\hline \hline 
	\end{tabular}
	\caption{The spin dependence of the pseudoscalar bilinear product $\overline{u}\gamma_5 v\,\overline{v}\gamma_5 u$ written in terms of momentum fractions and relative momenta.}
	\label{tab:spin_gamma5_qLR}
\end{table} 
\egroup
\subsection{Transverse and longitudinal integrals}
With the spin dependence of the Dirac bilinears in Eqs.~(\ref{eq:H_NJL_pi_matrix_wf},~\ref{eq:H_NJL_K_matrix_wf}) given by Tabs.~\ref{tab:spin_uuvv},~\ref{tab:spin_uvvu},~and~\ref{tab:spin_gamma5_qLR}, we proceed to evaluate the matrix elements of the NJL interactions in our basis representation. Explicitly, we calculate the corresponding cases where the matrix elements are defined in agreement with Eq.~\eqref{eq:Heff_scalar_uuvv_matrix_elements}. Doing so requires the transverse and longitudinal integrals explicitly calculated with the incoming and outgoing wavefunctions given in any configuration of the basis. 

Because the interactions defined by Eqs.~(\ref{eq:H_eff_NJL_pi},~\ref{eq:H_eff_NJL_K_ori}) contain no momentum transfer, the transverse integrals in Eq.~\eqref{eq:Heff_scalar_uuvv_matrix_elements} are reduced to the first few moments of the harmonic oscillator functions. Explicitly, the radial moments of $\phi_{nm}$ are given exactly as
\begin{align}
&\quad R_\mathrm{n}^M(b)\nonumber\\
& \equiv\int_{0}^{+\infty}\dfrac{qdq}{2\pi b}\sqrt{\dfrac{4\pi n!}{(n+M)!}}\left(\dfrac{q}{b}\right)^M e^{-q^2/(2b^2)}L_n^M(q^2/b^2)q^M\nonumber\\
& =\dfrac{2^M b^{M+1}\sqrt{n!}}{\sqrt{\pi (n+M)!}}\int_{0}^{+\infty}dy\,y^Me^{-y}L_n^M(2y)\nonumber\\%\nonumber\\
& =\dfrac{2^M b^{M+1}\sqrt{n!}}{\sqrt{\pi (n+M)!}}\sum_{m=0}^{n}\dfrac{\Gamma(n+M+1)(-2)^m}{\Gamma(n-m+1)\Gamma(m+1)}\nonumber\\
& =2^M b^{M+1}\sqrt{\dfrac{(n+M)!}{\pi n!}}(-1)^n,\label{eq:def_R_nm}
\end{align}
with $M=|m|$ and $q=|\overrightarrow{q}^\perp|$. For the second step of Eq.~\eqref{eq:def_R_nm}, we have applied the variable substitution ${q^2=2b^2y}$. Meanwhile, the angular parts of the transverse integrals are trivial to carry out. The transverse integrals needed are then given by 
\begin{subequations}\label{eq:phi_nm_moments}
	\begin{align}
	\int \dfrac{d\overrightarrow{q}^\perp}{(2\pi)^2}\, \phi_{nm}\left(\overrightarrow{q}^\perp\right)
	&=R_n^0(b)\delta_{m,0}
	=\dfrac{b}{\sqrt{\pi}}(-1)^n\delta_{m,0},
	\end{align}
	\begin{align}
	\int \dfrac{d\overrightarrow{q}^\perp}{(2\pi)^2} \,q^{\mathrm{R}}\,\phi_{nm}\left(\overrightarrow{q}^\perp\right)
	&=R_n^1(b)\delta_{m,-1}\nonumber\\
	& =2b^2\sqrt{\dfrac{n+1}{\pi}}(-1)^n\delta_{m,-1},
	\end{align}
	\begin{align}
	\int \dfrac{d\overrightarrow{q}^\perp}{(2\pi)^2} \,q^{\mathrm{L}}\,\phi_{nm}\left(\overrightarrow{q}^\perp\right) &
	=R_n^1(b)\delta_{m,1}\nonumber\\
	& =2b^2\sqrt{\dfrac{n+1}{\pi}}(-1)^n\delta_{m,1},
	\end{align}
	\begin{align}
	&\quad \int \dfrac{d\overrightarrow{q}^\perp}{(2\pi)^2} \,\left(q^{\mathrm{R}}\right)^2\,\phi_{nm}\left(\overrightarrow{q}^\perp\right)
	= R_n^2(b)\delta_{m,-2}\nonumber\\
	&=4b^3\sqrt{\dfrac{(n+1)(n+2)}{\pi}}(-1)^n\delta_{m,-2},
	\end{align}
	\begin{align}
	&\quad \int \dfrac{d\overrightarrow{q}^\perp}{(2\pi)^2} \,\left(q^{\mathrm{L}}\right)^2\,\phi_{nm}\left(\overrightarrow{q}^\perp\right)
	=R_n^2(b)\delta_{m,2}\nonumber\\
	&=4b^3\sqrt{\dfrac{(n+1)(n+2)}{\pi}}(-1)^n\delta_{m,2},
	\end{align}
	\begin{align}
	&\quad \int \dfrac{d\overrightarrow{q}^\perp}{(2\pi)^2} \,q^{\mathrm{L}} q^{\mathrm{R}}\,\phi_{nm}\left(\overrightarrow{q}^\perp\right)
	=R_n^2(b)\delta_{m,0}\nonumber\\
	&=4b^3\sqrt{\dfrac{(n+1)(n+2)}{\pi}}(-1)^n\delta_{m,0},
	\end{align}
\end{subequations}
while the integrals of $\phi_{nm}^{*}$ are  given by Eq.~\eqref{eq:phi_nm_moments} with $m\rightarrow -m$ for the corresponding Kronecker delta. 

With the transverse integrals reduced to Eq.~\eqref{eq:phi_nm_moments}, in order to calculate the matrix elements of the interactions in Eqs.~(\ref{eq:H_NJL_pi_matrix_wf},~\ref{eq:H_NJL_K_matrix_wf}), the remaining $x,~x'$ dependence in Tabs.~\ref{tab:spin_uuvv},~\ref{tab:spin_uvvu},~and~\ref{tab:spin_gamma5_qLR} contributes to the longitudinal integrals. These integrals can be further reduced to 
\begin{align}
&\quad L_l(a,b;\alpha,\beta) \nonumber\\
&\equiv \int_{0}^{1}\dfrac{dx}{4\pi}\,x^b(1-x)^a\chi_l(x;\alpha,\beta)\nonumber\\
& =\sqrt{\dfrac{2l+\alpha+\beta+1}{4\pi}}\sqrt{\dfrac{\Gamma(l+1)\Gamma(l+\alpha+\beta+1)}{\Gamma(l+\alpha+1)\Gamma(l+\beta+1)}}\nonumber\\
&\quad \times \int_{0}^{1}dx\,x^{\beta/2+b}(1-x)^{\alpha/2+a}\,P_l^{(\alpha,\beta)}(2x-1) \nonumber\\
&=\sqrt{\dfrac{2l+\alpha+\beta+1}{4\pi}}\sqrt{\dfrac{\Gamma(l+1)\Gamma(l+\alpha+\beta+1)}{\Gamma(l+\alpha+1)\Gamma(l+\beta+1)}}\nonumber\\
&\quad\times\sum_{m=0}^{l}
\begin{pmatrix}
l+\alpha \\ 
m
\end{pmatrix} 
\begin{pmatrix}
l+\beta \\ 
l-m
\end{pmatrix}(-1)^{l-m}\nonumber\\
&\quad\quad \times B\left(\dfrac{\beta}{2}+b+m+1,\dfrac{\alpha}{2}+a+l-m+1\right),\label{eq:def_L1_ab_alphabeta}
\end{align}
where $B(s,t)=\Gamma(s)\Gamma(t)/\Gamma(s+t)$ is the Euler Beta function. 

To evaluate $L_l(a,b;\alpha,\beta)$ numerically, we first rewrite Eq.~\eqref{eq:def_L1_ab_alphabeta} as 
\begin{equation}
L_l(a,b;\alpha,\beta)=\sqrt{\dfrac{2l+\alpha+\beta+1}{4\pi}}\sum_{m=0}^{l}C_{l,m}(a,b;\alpha,\beta),
\end{equation}
with
\begin{align}
&\quad C_{l,m}\nonumber\\
&\equiv\dfrac{(-1)^{l-m}\sqrt{\Gamma(l+1)\Gamma(l+\alpha+\beta+1)}}{\Gamma(m+1)\Gamma(l+\alpha-m+1)}\nonumber\\
&\quad \times \dfrac{\sqrt{\Gamma(l+\alpha+1)\Gamma(l+\beta+1)}}{\Gamma(l-m+1)\Gamma(\beta+m+1)}\nonumber\\
&\quad \times\dfrac{\Gamma(\beta/2+b+m+1)\Gamma(\alpha/2+a+l-m+1)}{\Gamma(\beta/2+b+\alpha/2+a+l+2)}.
\end{align}
We then obtain the following recurrence relations for $C_{l,m}$:
\begin{subequations}\label{eq:C_lm_recurrence}
	\begin{align}
	C_{0,0}& =\sqrt{\dfrac{\Gamma(\alpha+\beta+1)}{\Gamma(\alpha+1)\Gamma(\beta+1)}}\nonumber\\
	&\quad \times \dfrac{\Gamma(\beta/2+b+1)\Gamma(\alpha/2+a+1)}{\Gamma(\beta/2+b+\alpha/2+a+2)},\\[2mm]
	\dfrac{C_{l,0}}{C_{l-1,0}}& =-\sqrt{\dfrac{(l+\beta)(l+\alpha+\beta)}{l(l+\alpha)}}\nonumber\\
	&\quad \times \dfrac{\alpha/2+a+l}{\beta/2+b+\alpha/2+a+l+1}\quad\mathrm{for}\quad l\geq 1, 
	\end{align}
	\begin{align}
	\dfrac{C_{l,m}}{C_{l,m-1}}& =
	-\dfrac{(l+\alpha-m+1)(l-m+1)(\beta/2+b+m)}{m(\beta+m)(\alpha/2+a+l-m+1)} \nonumber\\
	&\quad \hspace{3cm}\mathrm{for}\quad l\geq m\geq 1.
	\end{align}
\end{subequations}
The longitudinal integral $L_l(a,b;\alpha,\beta)$ can then be calculated by first generating and then summing the following sequences:
\begin{equation*}
\begin{array}{l}
C_{0,0} \\ 
\downarrow \\ 
C_{1,0}+C_{1,1} \\ 
\downarrow \\ 
C_{2,0}+C_{2,1}+C_{2,2} \\ 
\downarrow \\ 
C_{3,0}+C_{3,1}+C_{3,2}+C_{3,3} \\ 
\downarrow \\ 
\dots\,\dots
\end{array},
\end{equation*}
using Eq.~\eqref{eq:C_lm_recurrence}.
\subsection{Matrix elements in the basis representation}
With both the longitudinal and transverse integrals known exactly, the explicit matrix elements are defined by Eq.~\eqref{eq:Heff_scalar_uuvv_matrix_elements}, and similarly the matrix elements $\big\langle n'm'l's_1's_2'\big\vert \overline{u} v\,\overline{v} u\big\vert nmls_1s_2\big\rangle$ and
$\big\langle n'm'l's_1's_2'\big\vert \overline{u}\gamma_5 v\,\overline{v}\gamma_5 u\big\vert nmls_1s_2\big\rangle$, are readily calculated. Results for these matrix elements are given by Tabs.~(\ref{tab:Heff_scalar_uuvv},~\ref{tab:Heff_scalar_uvvu},~and~\ref{tab:Heff_gamma5_uvvu}).
\bgroup
\def\arraystretch{1.5}
\begin{table*}\centering
	\begin{tabular}{c|c}
		\hline \hline  $s_1's_2's_2s_1$ & $\langle n'm'l's_1's_2'\vert \overline{u}u\,\overline{v} v\vert nmls_1s_2\rangle$ \\ 
		\hline $++++$ & $\begin{aligned} & (-1)^{n'+n+1}(b^2/\pi)\delta_{m',0}\delta_{m,0}\mathbf{m}\overline{\mathbf{m}} \big\{L'(1/2,1/2)L(-1/2,-1/2)\\
		&\quad +L'(-1/2,1/2)L(1/2,-1/2)+L'(1/2,-1/2)L(-1/2,1/2)+L'(-1/2,-1/2)L(1/2,1/2)\big\} \end{aligned}$ \\ 
		\hline $+++-$ & $\begin{aligned} &(-1)^{n'+n+1}(2b^3/\pi)\overline{\mathbf{m}}\Big\{\sqrt{n'+1}\delta_{m',-1}\delta_{m,0}\left[L'(1,0)L(-1/2,1/2)+L'(0,0)L(1/2,1/2)\right]\\
		&-\sqrt{n+1}\delta_{m',0}\delta_{m,1}\left[L'(1/2,1/2)L(0,0)+L'(-1/2,1/2)L(1,0)\right] \Big\} \end{aligned}$ \\ 
		\hline $++-+$ & $\begin{aligned} &(-1)^{n'+n}(2b^3/\pi)\mathbf{m}\Big\{\sqrt{n'+1}\delta_{m',-1}\delta_{m,0}\left[L'(0,1)L(1/2,-1/2)+L'(0,0)L(1/2,1/2)\right]\\
		&-\sqrt{n+1}\delta_{m',0}\delta_{m,1}\left[L'(1/2,1/2)L(0,0)+L'(1/2,-1/2)L(0,1)\right] \Big\} \end{aligned}$ \\ 
		\hline $++--$ & $\begin{aligned} & (-1)^{n'+n}(4b^4/\pi)\Big\{-\sqrt{(n'+1)(n+1)}\delta_{m',-1}\delta_{m,1}\left[L'(0,1)L(1,0)+L'(1,0)L(0,1)\right]\\
		&+\left[\sqrt{(n'+1)(n'+2)}\delta_{m',-2}\delta_{m,0}+\sqrt{(n+1)(n+2)}\delta_{m',0}\delta_{m,2}\right]L'(1/2,1/2)L(1/2,1/2) \Big\} \end{aligned} $ \\ 
		\hline $+-++$ & $\begin{aligned} & (-1)^{n'+n+1}(2b^3/\pi)\mathbf{m}\Big\{\sqrt{n'+1}\delta_{m',1}\delta_{m,0}\left[L'(0,1)L(1/2,-1/2)+L'(0,0)L(1/2,1/2)\right]\\
		&-\sqrt{n+1}\delta_{m',0}\delta_{m,-1}\left[L'(1/2,1/2)L(0,0)+L'(1/2,-1/2)L(0,1) \right] \Big\}\end{aligned}$ \\ 
		\hline $+-+-$ & $\begin{aligned} & (-1)^{n'+n}(4b^4/\pi)\Big\{ \sqrt{(n'+1)(n+1)}\left[\delta_{m',-1}\delta_{m,-1}L'(1,0)L(0,1)+\delta_{m',1}\delta_{m,1}L'(0,1)L(1,0)\right]\\
		& -\left[\sqrt{(n'+1)(n'+2)}+\sqrt{(n+1)(n+2)} \right]\delta_{m',0}\delta_{m,0}L'(1/2,1/2)L(1/2,1/2) \Big\}\end{aligned}$ \\ 
		\hline $+--+$ & $++++$ \\ 
		\hline $+---$ & $+++-$ \\ 
		\hline $-+++$ & $\begin{aligned} &(-1)^{n'+n                                                                                                                                                                                                                                                                                                                                                                                                                                                                                                                                                                                                   }(2b^3/\pi)\overline{\mathbf{m}}\Big\{\sqrt{n'+1}\delta_{m',1}\delta_{m,0}\left[L'(1,0)L(-1/2,1/2)+L'(0,0)L(1/2,1/2)\right]\\
		&-\sqrt{n+1}\delta_{m',0}\delta_{m,-1}\left[L'(1/2,1/2)L(0,0)+L'(-1/2,1/2)L(1,0)\right] \Big\} \end{aligned}$ \\ 
		\hline $-++-$ & $++++$ \\ 
		\hline $-+-+$ & $\begin{aligned} & (-1)^{n'+n}(4b^4/\pi)\Big\{ \sqrt{(n'+1)(n+1)}\left[\delta_{m',1}\delta_{m,1}L'(1,0)L(0,1)+\delta_{m',-1}\delta_{m,-1}L'(0,1)L(1,0)\right]\\
		& -\left[\sqrt{(n'+1)(n'+2)}+\sqrt{(n+1)(n+2)} \right]\delta_{m',0}\delta_{m,0}L'(1/2,1/2)L(1/2,1/2) \Big\}\end{aligned}$ \\ 
		\hline $-+--$ & $++-+$ \\ 
		\hline $--++$ & $\begin{aligned} & (-1)^{n'+n+1}(4b^4/\pi)\Big\{\sqrt{(n'+1)(n+1)}\delta_{m',1}\delta_{m,-1}\left[L'(0,1)L(1,0)+L'(1,0)L(0,1) \right]\\
		&-\left[\sqrt{(n'+1)(n'+2)}\delta_{m',2}\delta_{m,0}+\sqrt{(n+1)(n+2)}\delta_{m',0}\delta_{m,-2}\right]L'(1/2,1/2)L(1/2,1/2) \Big\} \end{aligned} $ \\ 
		\hline $--+-$ & $+-++$ \\ 
		\hline $---+$ & $-+++$ \\ 
		\hline $----$ & $++++$ \\ 
		\hline \hline 
	\end{tabular}
	\caption{The spin decomposition of the matrix elements defined by Eq.~\eqref{eq:Heff_scalar_uuvv_matrix_elements}. Here $L'(a,b)$ and $L(a,b)$ stand for $L_{l'}(a,b;\alpha,\beta)$ and $L_{l}(a,b;\alpha,\beta)$ respectively. Notice that the dependences on $l$ and $l'$ are implicit. The quark mass parameters have been written in boldface to distinguish them from the angular excitation number $m$.}\label{tab:Heff_scalar_uuvv}
\end{table*}
\egroup
\bgroup
\def\arraystretch{1.5}
\begin{table*}\centering
	\begin{tabular}{c|c}
		\hline \hline $s_1's_2's_2s_1$ & $\langle n'm'l's_1's_2'\vert \overline{u}v\,\overline{v} u\vert nmls_1s_2\rangle$ \\ 
		\hline $++++$ & $(-1)^{n'+n}(4b^4/\pi)\sqrt{(n'+1)(n+1)}\,\delta_{m',-1}\delta_{m,-1}\,L'(0,0)L(0,0)$ \\ 
		\hline $+++-$ & $(-1)^{n'+n}(2b^3/\pi)\sqrt{n'+1}\,\delta_{m',-1}\delta_{m,0}\,L'(0,0)[\mathbf{m}\,L(1/2,-1/2)-\overline{\mathbf{m}}\,L(-1/2,1/2)]$ \\ 
		\hline $++-+$ & $+++-$ \\ 
		\hline $++--$ & $(-1)^{n'+n+1}(4b^4/\pi)\sqrt{(n'+1)(n+1)}\,\delta_{m',-1}\delta_{m,1}\,L'(0,0)L(0,0)$ \\ 
		\hline $+-++$ & $(-1)^{n'+n}(2b^3/\pi)\sqrt{n+1}\,\delta_{m',0}\delta_{m,-1}\,[\mathbf{m}\,L'(1/2,-1/2)-\overline{\mathbf{m}}L'(-1/2,1/2)]L(0,0)$ \\ 
		\hline $+-+-$ & $\begin{aligned} & (-1)^{n'+n}(b^2/\pi)\,\delta_{m',0}\delta_{m,0}\,\big\{\mathbf{m}^2\,L'(1/2,-1/2)L(1/2,-1/2)\\
		& -\mathbf{m}\overline{\mathbf{m}}[L'(1/2,-1/2)L(-1/2,1/2)+L
		'(-1/2,1/2)L(1/2,-1/2)]+\overline{\mathbf{m}}^2\,L'(-1/2,1/2)L(-1/2,1/2)\big\} \end{aligned}$ \\ 
		\hline $+--+$ & $+-+-$ \\ 
		\hline $+---$ & $(-1)^{n'+n+1}(2b^3/\pi)\sqrt{n+1}\,\delta_{m',0}\delta_{m,1}\,[\mathbf{m}\,L'(1/2,-1/2)-\overline{\mathbf{m}}\,L'(-1/2,1/2)]L(0,0)$ \\ 
		\hline $-+++$ & $+-++$ \\ 
		\hline $-++-$ & $+-+-$ \\ 
		\hline $-+-+$ & $+-+-$ \\ 
		\hline $-+--$ & $+---$ \\ 
		\hline $--++$ & $(-1)^{n'+n+1}(4b^4/\pi)\sqrt{(n'+1)(n+1)}\,\delta_{m',1}\delta_{m,-1}\,L'(0,0)L(0,0) $ \\ 
		\hline $--+-$ & $(-1)^{n'+n+1}(2b^3/\pi)\sqrt{n'+1}\,\delta_{m',1}\delta_{m,0}\,L'(0,0)[\mathbf{m}\,L(1/2,-1/2)-\overline{\mathbf{m}}\,L(-1/2,1/2)]$ \\ 
		\hline $---+$ & $--+-$ \\ 
		\hline $----$ & $(-1)^{n'+n}(4b^4/\pi)\sqrt{(n'+1)(n+1)}\,\delta_{m',1}\delta_{m,1}\,L'(0,0)L(0,0)$ \\ 
		\hline \hline 
	\end{tabular}
	\caption{The spin decomposition of the matrix elements $\langle n'm'l's_1's_2'\vert \overline{u}v\,\overline{v} u\vert nmls_1s_2\rangle$. Here $L'(a,b)$ and $L(a,b)$ stand for $L_{l'}(a,b;\alpha,\beta)$ and $L_{l}(a,b;\alpha,\beta)$ respectively. Notice that the dependences on $l$ and $l'$ are implicit. The quark mass parameters have been written in boldface to distinguish them from the angular excitation number $m$.}\label{tab:Heff_scalar_uvvu}
\end{table*}
\egroup
\bgroup
\def\arraystretch{1.5}
\begin{table*}\centering
	\begin{tabular}{c|c}
		\hline \hline $s_1's_2's_2s_1$ & $\langle n'm'l's_1's_2'\vert \overline{u}\gamma^5v\,\overline{v}\gamma^5 u\vert nmls_1s_2\rangle$ \\ 
		\hline $++++$ & $(-1)^{n'+n+1}(4b^4/\pi)\sqrt{(n'+1)(n+1)}\delta_{m',-1}\delta_{m,-1}L'(0,0)L(0,0)$ \\ 
		\hline $+++-$ & $(-1)^{n'+n+1}(2b^3/\pi)\sqrt{n'+1}\delta_{m',-1\delta_{m,0}}L'(0,0)[\mathbf{m}L(1/2,-1/2)+\overline{\mathbf{m}}L(-1/2,1/2) ]$ \\ 
		\hline $++-+$ & $(-1)\,+++-$ \\ 
		\hline $++--$ & $(-1)^{n'+n+1}(4b^4/\pi)\sqrt{(n'+1)(n+1)}\delta_{m',-1}\delta_{m,1}L(0,0)L'(0,0)$ \\ 
		\hline $+-++$ & $(-1)^{n'+n}(2b^3/\pi)\sqrt{n+1}\delta_{m',0}\delta_{m,-1}[\mathbf{m}L'(1/2,-1/2)+\overline{\mathbf{m}}L'(-1/2,1/2)]L(0,0)$ \\ 
		\hline $+-+-$ & $(-1)^{n'+n}(b^2/\pi)\delta_{m',0}\delta_{m,0}[\overline{\mathbf{m}}L'(-1/2,1/2)+\mathbf{m}L'(1/2,-1/2)][\overline{\mathbf{m}}L(-1/2,1/2)+\mathbf{m}L(1/2,-1/2)]$ \\ 
		\hline $+--+$ & $(-1)\,+-+-$ \\ 
		\hline $+---$ & $(-1)^{n'+n}(2b^3/\pi)\sqrt{n+1}\delta_{m',0}\delta_{m,1}[\overline{\mathbf{m}}L'(-1/2,1/2)+\mathbf{m}L'(1/2,-1/2)]L(0,0)$ \\ 
		\hline $-+++$ & $(-1)\,+-++$ \\ 
		\hline $-++-$ & $(-1)\,+-+-$ \\ 
		\hline $-+-+$ & $+-+-$ \\ 
		\hline $-+--$ & $(-1)\,+---$ \\ 
		\hline $--++$ & $(-1)^{n'+n+1}(4b^4/\pi)\sqrt{(n'+1)(n+1)}\delta_{m',1}\delta_{m,-1} L'(0,0)L(0,0)$ \\ 
		\hline $--+-$ & $(-1)^{n'+n+1}(2b^3/\pi)\sqrt{n'+1}\delta_{m',1}\delta_{m,0}L'(0,0)[\overline{\mathbf{m}}L(-1/2,1/2)+\mathbf{m}L(1/2,-1/2)]$ \\ 
		\hline $---+$ & $(-1)\,--+-$ \\ 
		\hline $----$ & $(-1)^{n'+n+1}(4b^4/\pi)\sqrt{(n'+1)(n+1)}\delta_{m',1}\delta_{m,1}L'(0,0)L(0,0)$ \\ 
		\hline \hline 
	\end{tabular}
	\caption{Spin dependences of the matrix elements $\langle n'm'l's_1's_2'\vert \overline{u}\gamma^5v\,\overline{v}\gamma^5 u\vert nmls_1s_2\rangle$. Here $L'(a,b)$ and $L(a,b)$ stand for $L_{l'}(a,b;\alpha,\beta)$ and $L_{l}(a,b;\alpha,\beta)$ defined by Eq.~\eqref{eq:def_L1_ab_alphabeta} respectively. Notice that the dependences on $l$ and $l'$ are implicit. The quark mass parameters have been written in the boldface to distinguish them from the angular excitation number $m$.}\label{tab:Heff_gamma5_uvvu}
\end{table*}
\egroup
	\bibliography{BLFQ_NJL.bib}
\end{document}